\documentclass[pdflatex,bst/sn-mathphys-num]{sn-jnl}

\usepackage{graphicx}%
\usepackage{multirow}%
\usepackage{amsmath,amssymb,amsfonts}%
\usepackage{amsthm}%

\usepackage{mathrsfs}%
\usepackage[title]{appendix}%
\usepackage{xcolor}%
\usepackage{textcomp}%
\usepackage{manyfoot}%
\usepackage{booktabs}%
\usepackage{algorithm}%
\usepackage{algorithmicx}%
\usepackage{algpseudocode}%
\usepackage{listings}%
\usepackage{natbib}
\usepackage{subfig}
\usepackage{mathtools}
\DeclareMathOperator*{\argminA}{arg\,min}

\theoremstyle{thmstyleone}%
\theoremstyle{thmstyletwo}%
\newtheorem{remark}{Remark}%

\theoremstyle{thmstylethree}%

\begin{document}

\title[A Stable Lasso]{A Stable Lasso}


\author[1]{\fnm{Mahdi} \sur{Nouraie}}

\author[1]{\fnm{Houying} \sur{Zhu}}

\author*[1,2]{\fnm{Samuel} \sur{Muller}}\email{samuel.muller@mq.edu.au}

\affil[1]{\orgdiv{School of Mathematical and Physical Sciences}, \orgname{Macquarie University}, \orgaddress{\city{Sydney}, \postcode{2109}, \state{NSW}, \country{Australia}}}
\affil[2]{\orgdiv{School of Mathematics and
Statistics}, \orgname{The University of Sydney}, \orgaddress{\city{Sydney}, \postcode{2006}, \state{NSW}, \country{Australia}}}


\abstract{The Lasso has been widely used as a method for variable selection, valued for its simplicity and empirical performance. However, Lasso's selection stability deteriorates in the presence of correlated predictors. Several approaches have been developed to mitigate this limitation. In this paper, we provide a brief review of existing approaches, highlighting their limitations. We then propose a technique to improve the selection stability of Lasso by integrating a weighting scheme into the Lasso penalty function, where the weights are defined as an increasing function of a correlation-adjusted ranking that reflects the predictive power of predictors. Empirical evaluations on both simulated and real-world datasets demonstrate the efficacy of the proposed method. Additional numerical results demonstrate the effectiveness of the proposed approach in stabilizing other regularization-based selection methods, indicating its potential as a general-purpose solution.}




\maketitle

\section{Introduction}\label{s1}
The Least Absolute Shrinkage and Selection Operator \citep[Lasso;][]{tibshirani1996regression} has been widely used as a regularization-based variable selection method. However, it is well-established that the Lasso becomes instable in the presence of correlated predictor variables \citep{10.1093/bioinformatics/bty750}. Specifically, when an irrelevant variable is highly correlated with relevant variables, the Lasso may be unable to distinguish between them, regardless of the amount of data available or the degree of regularization applied \citep{a1294e48-ce43-314c-867d-a032d7baf484, zhao2006model}. In this paper, we briefly review existing approaches aimed at addressing this issue, highlighting their limitations. We subsequently propose a methodology that improves the selection stability of the Lasso without increasing the optimization cost, altering the original data space, or requiring prior knowledge of the underlying grouping structure among predictors, addressing key limitations of existing methods. Our methodology relies on weighting predictors within the Lasso optimization process according to a correlation-adjusted ranking that reflects their predictive power.

We consider a dataset $\mathcal{D} = \{(\boldsymbol{x}^\top_{i}, y_{i})\}_{i=1}^{n}$ where each element of $\mathcal{D}$ consists of a univariate response $y_{i} \in \mathbb{R}$ and a $p$-dimensional fixed vector of predictor variables $\boldsymbol{x}^\top_{i} \in \mathbb{R}^{p}$. Linear regression is formally defined as $\boldsymbol{Y} =  X\boldsymbol{\beta} + \boldsymbol{\varepsilon}$, where $\boldsymbol{Y}$ denotes the $n-$vector of the response variable, $X \in \operatorname{mat}(n,p)$ denotes the fixed design matrix,  $\boldsymbol{\beta}$ denotes the $p-$vector of regression coefficients, and $\boldsymbol{\varepsilon}$ denotes the $n-$vector of the random errors. The intercept term is omitted here because, without loss of generality,  we assume that the response vector is centered; that is, $\sum_{i = 1}^{n}y_{i} = 0$. We also assume that the columns of the design matrix $X$ are centred and scaled by their means and standard deviations. Following the basic assumptions of the linear regression, it is assumed that $\mathbb{E}(\boldsymbol{Y}|X)$ is linear in the coefficients, $\mathbb{E}(\boldsymbol{\varepsilon}|X) = 0$, and $\text{Var}(\boldsymbol{\varepsilon}|X) = \sigma^2\mathbb{I}_{n}$ where $\sigma^2 < \infty$ and $\mathbb{I}_{n} \in \operatorname{mat}(n \times n)$ denotes the identity matrix.

The Lasso performs variable selection by augmenting the least-squares loss with an $\ell_1-$norm penalty on the regression coefficients, thereby encouraging sparsity in the estimated model. The Lasso regression coefficients are estimated through 
\begin{equation}\label{eqn-Lasso}
\boldsymbol{\hat{\beta}}(\lambda) = \argminA_{\boldsymbol{\beta} \in \mathbb{R}^{p}} \left(\|\boldsymbol{Y} - X\boldsymbol{\beta}\|_{2}^{2} + \lambda \|\boldsymbol{\beta}\|_{1}\right),
\end{equation}
where $\lambda \in \mathbb{R}^{+}$ denotes the Lasso regularization parameter.
Numerous extensions of the Lasso have been proposed since its inception (see, e.g., \citet{zou2005regularization}, \citet{tibshirani2005sparsity}, \citet{zou2006adaptive}, \citet{yuan2006model}, \citet{park2008bayesian}, \citet{pmlr-v9-lorbert10b}, \citet{NIPS2011_33ceb07b}, \citet{simon2013sparse}, \citet{Chen_Ding_Luo_Xie_2013}, \citet{NIPS2014_43feaeee}, \citet{pmlr-v84-takada18a},
\citet{10.1093/bioinformatics/bty750}, and 
\citet{craig2024pretraining}), highlighting its popularity and effectiveness.

The stability of a variable selection method pertains to its capacity to consistently identify the same variables across different training sets sampled from the same underlying distribution \citep{kalousis2007stability}. This property serves as an indicator of the reproducibility and generalizability of the selection results and has garnered significant attention in recent years  (see, e.g., \citet{https://doi.org/10.1111/j.1751-5823.2010.00108.x}, \citet{meinshausen2010stability}, \citet{10.1093/bioinformatics/bty750}, and \citet{JMLR:v25:23-0536}). The presence of correlated predictor variables in $X$ is known to adversely affect the stability of variable selection performed by Lasso \citep{meinshausen2010stability, 10.1093/bioinformatics/bty750, faletto2022cluster, nouraie2024selection}.

Following \citet{pmlr-v84-takada18a}, variable selection methods can be broadly categorized into two classes. The first, known as `grouping selection', seeks to retain all correlated variables that collectively provide predictive information for the response. The second, referred to as `exclusive selection', retains only a subset of such correlated variables, based on the premise that including all may lead to redundancy. In this paper, we focus on exclusive selection methods. In this research thread, \citet{Chen_Ding_Luo_Xie_2013} introduced the Uncorrelated Lasso, which was later refined by \citet{pmlr-v84-takada18a} into the Independently Interpretable Lasso (IILasso). IILasso addresses the challenges posed by correlated variables by incorporating an additional penalty term into the Lasso objective function. This penalty is designed to discourage the simultaneous selection of correlated variables, encouraging sparse solutions within correlated groups. To the best of our knowledge, IILasso represents the current state-of-the-art in exclusive selection. Although effective, its optimization requires the full covariance matrix of the predictor variables, making it computationally expensive in high-dimensional settings, where $p \gg n$. In this paper, we propose an alternative strategy whose optimization reduces to the standard Lasso, enhanced by variable-specific weights derived from a correlation-adjusted ranking based on the predictive power of predictors.

Certain methods, such as the Sparse Group Lasso \citep[SGL;][]{simon2013sparse}, have been developed to encourage sparsity both between and within groups of correlated variables. However, these approaches typically require prior knowledge of the underlying grouping structure among predictors, which distinguishes them from the focus of this paper. In addition, similar to IILasso, these methods are generally more computationally intensive owing to the complexity of their objective functions. On the other hand, Stability Selection was introduced as a general framework to improve the stability of variable selection methods by leveraging random sub-samples of original data \citep{meinshausen2010stability, shah2013variable}. Although this approach has shown considerable success, it has been demonstrated that it does not resolve the instability of Lasso in the presence of correlated predictors, due to the vote-splitting effect \citep{meinshausen2010stability, faletto2022cluster}. In this paper, we use Stability Selection to assess the selection stability and to calibrate the regularization parameter $\lambda$, following the methodology proposed by \citet{nouraie2024selection}, as outlined in Section \ref{s2}.

\citet{nouraie2025stability} proposed an alternative strategy for handling correlated variables by decorrelating them prior to selection. Their method, as a modified version of the classic Gram-Schmidt orthogonalization \citep{BJORCK1994297}, applies selection algorithm to an orthonormal surrogate rather than the original data. While effective, this approach alters the original variable space. In this work, we explore a complementary direction that aims to improve selection stability of Lasso while preserving the original design matrix.

\citet{zou2006adaptive} proposed the Adaptive Lasso that incorporates variable-specific weights into the Lasso penalty, with the weights derived from initial estimates of the regression coefficients. In high-dimensional settings, obtaining consistent initial estimates is challenging. \citet{bhl2011} suggested using Lasso estimates as a basis for the initial weights; however, given the known instability of Lasso, propagating its estimates into the weighting scheme may reinforce the initial instable results. \citet{a2ce9dde-7f22-3541-a877-07b4efae8445} showed that when relevant and irrelevant variables are only weakly correlated, marginal univariate regression can provide suitable initial estimates. Nevertheless, this assumption rarely holds in real-world datasets, where strong correlations among predictors are common. 

\citet{meinshausen2010stability} introduced the Randomized Lasso, a method conceptually related to the Adaptive Lasso, with the primary distinction that variable weights are assigned randomly. As described by \citet{meinshausen2010stability}, the Randomized Lasso modifies the standard Lasso by weighting each predictor in the objective function by $1/w_j$, where $\Pr(w_j = \alpha) \in (0,1)$, and $w_j = 1$ otherwise for $j = 1,2,\dots, p$. They demonstrated that this randomization yields asymptotic consistency within the Stability Selection framework, assuming sparse eigenvalues, a condition weaker than the irrepresentable condition \citep{zhao2006model, meinshausen2006high} which guarantees consistent variable selection for the Lasso. Although the Randomized Lasso achieves asymptotic consistency, we contend that its random weighting scheme may introduce additional variability and potentially undermine the accuracy of the results. A numerical comparison between our method and the Randomized Lasso is provided in Section~\ref{s3}. \citet{courtois2021new} also proposed two variants of the Adaptive Lasso that aim to improve false discovery control and sensitivity. While Adaptive Lasso shares a conceptual similarity with our approach, the way we assign the weights is notably different. To the best of our knowledge, no existing variant of the Adaptive Lasso accounts for the correlation structure among predictors and incorporates their predictive relevance rankings to inform variable-specific weights. Other related methods are discussed in Section~\ref{s4}.

\citet{10.1111/j.1467-9868.2008.00674.x} introduced the Sure Independence Screening (SIS) procedure, which reduces the dimensionality of high-dimensional data by ranking variables based on their marginal correlations with the response variable. This approach asymptotically retains relevant variables with overwhelming probability, a desirable property known as the `sure screening property'. According to \citet{10.1111/j.1467-9868.2008.00674.x}, a sufficient condition for SIS to possess this property is that the marginal correlations of relevant predictors must be bounded away from zero. However, this condition is often violated in practice due to the presence of correlation among predictors \citep{10.1111/rssb.12127}. We highlight the similarity between this issue and that discussed above concerning the use of marginal univariate regression coefficients in the Adaptive Lasso. Specifically, when the Adaptive Lasso uses the marginal coefficients to weight predictors, it ranks variables based on their individual predictive strength and applies a penalty that is implicitly an increasing function of their rank, without accounting for the underlying correlation structure among predictors. As mentioned above, this approach is reliable only under conditions of weak correlation.

\citet{10.1111/rssb.12127} introduced the Ridge High-dimensional Ordinary Least Squares Projection (Ridge-HOLP; \citealp{10.1111/rssb.12127}), a method that ranks variables by simultaneously accounting for their predictive power and correlation structure. This is achieved by removing the random diagonal matrix $D$ that arises in the singular value decomposition of $X$ \citep{10.1111/rssb.12127}. They demonstrated that when $p > n$, the Moore-Penrose inverse of $X$ can be used to construct a screening-consistent procedure; that is, when selecting a model of the same size as the true model, the method asymptotically identifies the correct set of relevant variables. Ridge-HOLP employs a diagonally dominant projection matrix $X^\top (X X^\top)^{-1} X$, such that multiplying this matrix by $\boldsymbol{\beta}$ is likely to preserve the rank order of the entries in $\boldsymbol{\beta}$ \citep{10.1111/rssb.12127}.

According to \citet{10.1111/rssb.12127} and \citet{10.1111/rssb.12427}, a critical condition for the screening consistency of Ridge-HOLP is $\kappa(\Sigma) \leq c n^{\tau}$, where $c > 0$ and $0 \leq \tau < 1/7.5$, $\Sigma$ denotes the covariance matrix of $X$, $\kappa(\Sigma) = \lambda_{\max}(\Sigma) / \lambda_{\min}(\Sigma)$ denotes the condition number of $\Sigma$, and $\lambda_{\max}(\Sigma)$ and $\lambda_{\min}(\Sigma)$ denote the largest and smallest eigenvalues of $\Sigma$ respectively. One instance in which this condition is met is when the groups of correlated variables are mutually independent, each group contains finitely many variables, and the variables within each group exhibit a compound symmetric structure with a correlation value less than one. The assumptions required for the screening consistency of Ridge-HOLP are stated in Theorem 3 of \citet{10.1111/rssb.12127} and \citet{10.1111/rssb.12427}. Recently, \citet{AirHOLP} introduced Air-HOLP, a data-adaptive extension of Ridge-HOLP, and demonstrated through extensive simulation studies that it performs well in the presence of correlated predictors.

In this paper, we propose to first rank the predictors using a correlation-adjusted ranking method based on \citet{10.1111/rssb.12127}, and then leverage these ranks to weight the predictors in the Lasso objective function, with the aim of enhancing selection stability. Consequently, our method can be viewed as a more stable variant of the Lasso and a generalized form of the Adaptive Lasso. The numerical results are highly promising compared to existing methods, particularly those with complex objective functions that result in substantially longer running times.

Unlike existing weighted penalization approaches that primarily focus on estimation consistency or support recovery, the proposed Stable Lasso is motivated by the goal of improving variable selection stability in correlated high-dimensional settings. The method combines two methodological paradigms that are traditionally studied separately by leveraging correlation-aware variable screening information from Air-HOLP to guide downstream variable selection through predictor-specific penalization. Compared with methods that explicitly model correlation structure at the optimization level, such as Uncorrelated Lasso or IILasso, our approach preserves the computational simplicity of standard Lasso while still accounting for dependence among predictors. Furthermore, unlike resampling-based approaches such as Stability Selection that improve reproducibility through aggregation, our method aims to improve stability directly by construction through a deterministic penalty structure. Importantly, as demonstrated in our numerical studies, the proposed weighting framework also improves the stability of other variable selection procedures beyond Lasso, suggesting that the broader contribution of this work extends beyond a single algorithmic modification and may represent a general technique for improving stability while preserving interpretability in high-dimensional correlated inference.

The rest of this paper is organized as follows. Section \ref{s2} outlines the proposed methodology.  Section \ref{s3} describes the real and synthetic datasets used in this paper and presents the corresponding numerical results. Section \ref{s4} reviews some related works and highlights some encouraging directions for future research.

\section{Methodology}\label{s2}
In this section, we propose a methodology to achieve stable variable selection using the Lasso, without increasing the computational cost of the optimization process, altering the original data, or requiring prior knowledge of the underlying grouping structure among predictor variables. The central idea is to assign weights to variables based on a correlation-adjusted ranking that reflects their predictive relevance. By correlation-adjusted, we mean that a relevant variable correlated with irrelevant ones is assigned a higher rank. These ranks are then transformed into variable-specific penalty factors and incorporated into the Lasso objective function. This weighting scheme provides the model with prior information, enabling it to prioritize more promising variables within groups of correlated predictors. As a result, the method yields more stable selection results by mitigating the random selection among correlated variables by the Lasso, as discussed in \citet{meinshausen2010stability}, \citet{10.1093/bioinformatics/bty750}, and \citet{faletto2022cluster}. However, the overall selection accuracy remains dependent on the output of the initial ranking.

Since this paper focuses on correlated variables and seeks to prioritize the relevant ones among them, it is important to employ a ranking method that assigns higher ranks to relevant variables within correlated groups. As noted by \citet{nouraie2025stability}, this requirement is weaker than screening consistency, as introduced in Section \ref{s1}, since it does not demand that all relevant variables to be assigned higher ranks than all irrelevant ones, but only that each relevant variable is ranked higher than the irrelevant variables with which it is correlated. We refer to this property as `partial screening consistency'.
Accordingly, Ridge-HOLP, which satisfies asymptotic screening consistency, serves as a suitable choice. 

In this paper, we employ Air-HOLP, a data-adaptive extension of Ridge-HOLP that iteratively optimizes the ridge parameter $r$, to construct the initial ranking of predictors. Assuming $s \asymp n^\nu$ for a $\nu < 1$ while $s$ denotes the size of the true model, we set the screening threshold $d = n / \log(n)$ which is a common choice in the screening literature \citep{AirHOLP}. While alternative choices for the screening threshold exist and have been proposed, a comprehensive comparison of them is beyond the scope of this paper. Although this paper focuses on the HOLP family of ranking algorithms, the proposed framework is sufficiently general to accommodate other correlation-adjusted screening and ranking procedures, including that of \citet{wang2025ridge}. Evaluating and comparing these alternatives falls beyond the scope of the present work. The chosen ranking method is expected to order variables such that those with greater predictive relevance to the response are assigned higher ranks.

Air-HOLP inherits the screening consistency of Ridge-HOLP by restricting the admissible domain of the ridge parameter $r$. Specifically, Theorem 3 in \citet{10.1111/rssb.12127} and \citet{10.1111/rssb.12427} establishes that, for Ridge-HOLP to attain the sure screening property, $r$ must satisfy $r = o(n^{1 - \frac{5}{2}\tau - \kappa})$, subject to the constraint $1 - 7.5\tau - 2\kappa - \nu > 0$, with $\tau, \kappa, \nu > 0$. Notably, any choice of $r = c n^a, 0 \leq a \leq 0.5, c \geq 0$, meets this condition. Accordingly, for Air-HOLP, to guarantee the sure screening property, the tuning parameter is restricted to the interval $[0, c\sqrt{n}]$, with $c = 1000$ employed unless otherwise specified \citep{AirHOLP}.

Applying Air-HOLP to $X$ and $\boldsymbol{Y}$ and obtaining the ranking of predictors, we define
\begin{equation}\label{eqn-weights}
w(r_j) =
1 - \frac{1}{r_j}; \quad j = 1,2,\dots, p,
\end{equation}
where $w_j = w(r_j)$ denotes the weight (penalty factor) assigned to the $j$th predictor, and $r_j$ denotes the rank assigned to the $j$th predictor using Air-HOLP. Ranks obtained by Air-HOLP represent the relative predictive relevance of predictors, with higher ranks indicating greater relevance.

To clarify the structural motivation behind the proposed weighting function, we explicitly relate its construction to the imposed rank-based desiderata. Let $r_j \in \{1,\dots,p\}$ denote the rank of a variable, where $r_j=1$ corresponds to the most relevant variable and increasing $r_j$ corresponds to decreasing relevance. The weighting function $w(r_j)$ is required to satisfy: (i) monotonicity, ensuring $w'(r_j)\ge 0$ so that lower-ranked variables receive larger penalty weights; (ii) boundedness, $w(r_j)\in[0,1]$; and (iii) asymptotic compatibility, $\lim_{r_j\to\infty} w(r_j)=1$.

To explicitly characterize deviation from the baseline Lasso penalty, we parameterize the weighting function as
\[
w(r_j)=1-f(r_j),
\]
where $f(r_j)\ge 0$ is a decreasing function capturing the rank-induced deviation from the baseline penalty. Since the basic structural requirements (monotonicity, boundedness, and asymptotic compatibility) do not uniquely determine $w(r_j)$, an additional modeling assumption is required.

We adopt a constant-elasticity specification for $f(r_j)$, which imposes proportional (relative) decay of rank effects. This modeling assumption yields a tractable family of admissible weighting functions indexed by a decay parameter $c>0$. This parameter governs the rate at which rank influence is attenuated and is not identified by the structural conditions alone.

Specifically, imposing this constant-elasticity condition yields a power-law form for the deviation from the baseline penalty, and consequently a parametric family of weighting functions of the form
\[
w(r_j)=1-Kr_{j}^{-c}.
\]
In Section 3, we compare this modeling choice with several alternative monotone weighting functions.

Although both $K$ and $c$ appear as parameters in this general form, they play fundamentally different roles. The parameter $c$ controls the relative decay rate of the rank effect and determines how rapidly the weighting function approaches the standard Lasso penalty as the rank increases. In contrast, $K$ acts primarily as a scale parameter that shifts the overall magnitude of the rank-based adjustment. Without an additional normalization, these two parameters may introduce unnecessary flexibility, as changes in $K$ can partially compensate for changes in $c$, making their individual effects difficult to interpret. 

To obtain an identifiable and interpretable weighting scheme, we impose the natural normalization $w(1)=0$, requiring the highest-ranked variable to receive no rank-based penalty adjustment. This condition uniquely determines $K=1$, leaving $c$ as the sole parameter controlling the decay of rank influence. Therefore,
\[
w(r_j)=1-r_{j}^{-c}.
\]

Within this family, smaller values of $c$ correspond to slower decay of rank influence and a more gradual transition to the standard Lasso penalty, whereas larger values concentrate the adjustment more strongly on the highest-ranked variables. To ensure robustness, we report a sensitivity analysis over different values of $c$ in Section 3, showing that the main conclusions are stable across a wide range of decay rates.

By incorporating variable-specific weights from Equation~\eqref{eqn-weights}, Equation~\eqref{eqn-Lasso} can be extended to
\begin{equation}\label{eqn-SLasso}
\boldsymbol{\hat{\beta}}(\lambda,  \boldsymbol{w}) = \argminA_{\boldsymbol{\beta} \in \mathbb{R}^{p}} \left(\|\boldsymbol{Y} - X\boldsymbol{\beta}\|_{2}^{2} + \lambda \|\boldsymbol{w} \odot \boldsymbol{\beta}\|_{1}\right),
\end{equation}
where $\boldsymbol{w}$ denotes the $p$-vector of $w_j$ values, and $\odot$ denotes the element-wise (Hadamard) product operator. The variable with the highest rank receives no additional penalty beyond the standard least-squares quadratic loss. As the rank increases, the corresponding weight gradually approaches $1$, matching the standard Lasso penalty. Hereafter, we refer to the model introduced in Equation~\eqref{eqn-SLasso} as the `Stable Lasso'.

Rather than relying solely on the asymptotic screening consistency of the screening method and discarding variables excluded from the screened set, we incorporate a rank-based weighting to guide the Lasso. The slope of the weight function introduced in Equation~\eqref{eqn-weights} decreases with increasing rank, implying that a unit increase in rank leads to a more substantial change in the weight for higher-ranked variables than for lower-ranked ones. As a result, the weights become less sensitive to rank changes among lower-ranked variables, effectively applying the standard Lasso to them. This behavior is reminiscent of certain methods, such as those proposed by \citet{Fan01122001} and \citet{MCP}, which aim to reduce the bias of the Lasso by being more lenient towards variables with larger coefficient magnitudes.

In terms of computational cost, \citet{AirHOLP} noted that in the high-dimensional setting, Ridge-HOLP is more computationally efficient than the standard ridge regression \citep{tikhonov1963solution, hoerl1970ridge}, and showed that while Air-HOLP uses an iterative process to update the initial tuning parameter, it remains considerably faster than the standard ridge regression in the high-dimensional setting.

We embed the Stable Lasso within Stability Selection to assess its selection stability across random sub-samples and to tune the hyper-parameter $\lambda$, as will be explained shortly. In this paper, the binary selection outcomes obtained from Stability Selection with the Stable Lasso for a fixed $\lambda$ are represented by the matrix $M(\lambda) \in \operatorname{mat}(B \times p)$, where $B$ denotes the number of sub-samples. The stability measure, $\hat{\Phi}(M(\lambda))$, is then computed as defined in \citet{nogueira2018stability}, corresponding to a scaled average of the column-wise variances of $M(\lambda)$. The stability estimator proposed by \citet{nogueira2018stability} is defined as
\begin{equation}\label{eqn-phi}
    \hat{\Phi}(M(\lambda)) = 1 - \frac{\frac{1}{p}\sum_{j=1}^{p} s_{j}^{2}}{\frac{q(\lambda)}{p}\left(1 - \frac{q(\lambda)}{p}\right)};\quad \lambda \in \Lambda,
\end{equation}
where $s_{j}^{2}$ denotes the unbiased sample variance of the binary selection statuses of the $j$th variable, while $q(\lambda)$ denotes the average number of variables selected under the regularization parameter $\lambda$. The stability estimator $\hat{\Phi}(\cdot)$ is bounded by $\left[-\frac{1}{B-1}, 1\right]$ \citep{nogueira2018stability}; the larger the $\hat{\Phi}(M(\lambda))$, the more stable is $M(\lambda)$.

\begin{remark}
Consider Stability Selection with the Lasso for a fixed $\lambda$ value. Then, employing a fixed non-uniform weighting scheme as penalty factors for the Lasso, applied consistently across all random sub-samples and recorded in $M'(\lambda)$, is expected to yield greater stability than applying the standard equally weighted Lasso and recording in $M''(\lambda)$.
\label{remark1}
\end{remark}

It is important to note that Remark 1 concerns the algorithmic stability of the selection procedure itself rather than the statistical accuracy of the resulting selected model. In particular, the remark does not imply that introducing arbitrary fixed non-uniform weights necessarily improves variable selection performance or guarantees identification of the truly relevant predictors. Poorly chosen weighting schemes may indeed stabilize the repeated selection of irrelevant variables. The purpose of this remark is instead to highlight a general mechanism by which deterministic non-uniform penalization reduces variability in selection outcomes by systematically resolving arbitrary competition among correlated predictors. The separate question of whether increased stability is accompanied by improved selection accuracy depends critically on the quality of the weighting mechanism used to construct the penalty factors.

The conclusion in Remark \ref{remark1} follows directly from the influence of weighting on the Lasso estimator. Random selection among correlated predictors is a well-known source of instability in the Lasso \citep{10.1093/bioinformatics/bty750}. By consistently applying a fixed, non-uniform weighting scheme across all sub-samples in Stability Selection, a systematic preference among predictors is enforced, independent of the value of $\lambda$ or the specific sub-sample. This consistency tends to align selection outcomes across sub-samples, reducing instability arising from arbitrary competition among correlated predictors and thereby potentially improving algorithmic stability in such settings.

As is evident, Remark \ref{remark1} is not restricted to the Stable Lasso: applying a fixed non-uniform weighting scheme in the Lasso generally enhances stability. However, some weighted Lasso variants employ nearly uniform weights. For instance, the Randomized Lasso uses only two weights, $1/\alpha$ and $1$, while in the Adaptive Lasso, when initial estimates are obtained from the standard Lasso, most regression coefficients are zero. Nevertheless, as will be demonstrated in Section \ref{s3}, even these methods often exhibit greater stability than the standard Lasso, which employs uniform weights. It should be noted, however, that stability alone is insufficient; selection accuracy must also be taken into account.

In addition, Remark \ref{remark1} is not restricted to the high-dimensional setting. Therefore, following the findings of \citet{nouraie2025stability}, we conclude that selection stability is also a concern in low-dimensional contexts. In Section \ref{s3}, we evaluate the performance of the Stable Lasso in a low-dimensional setting to examine its effectiveness under such conditions.

\begin{remark}
Under the assumptions of Theorem 3 in \citet{10.1111/rssb.12127} and \citet{10.1111/rssb.12427}, for an appropriately chosen $\lambda$, the Stable Lasso leverages the screening consistency property of Ridge-HOLP to assign lower penalties to highly ranked variables and larger penalties to low-ranked variables. Consequently, relevant predictors are expected to be selected more frequently, while irrelevant predictors face a higher effective threshold for entering the model, potentially reducing false discoveries.
\label{remark2}
\end{remark}


Theorem 3 in \citet{10.1111/rssb.12127} establishes the asymptotic screening consistency of Ridge-HOLP, showing that, with increasing sample size, all relevant variables are ranked above irrelevant ones with high probability. The Stable Lasso leverages this ranking by applying variable-specific penalty weights, assigning smaller penalties to higher-ranked, potentially relevant variables. Under the sufficient conditions stated in Remark \ref{remark2}, this mechanism promotes the selection of all relevant variables. Notably, the necessary requirement is partial screening consistency, which is substantially weaker than full screening consistency.

For the optimization of Equation~\eqref{eqn-SLasso}, we employ a coordinate descent algorithm, as recommended by \citet{friedman2010regularization}, incorporating the weight vector $\boldsymbol{w}$ as a penalty factor within the Lasso objective function.

\subsection*{Hyper-Parameter Tuning}
Equation~\eqref{eqn-SLasso} includes the hyper-parameter $\lambda \in \Lambda$ that governs the overall level of shrinkage applied during estimation. \citet{nouraie2024selection} proposed tuning $\lambda$ with respect to both selection stability and prediction accuracy, yielding a Pareto optimal solution \citep{pareto1896cours} on the stability-accuracy Pareto front, as conceptualized by \citet{nogueira2018stability}, and in line with \citet{10.5555/2567709.2567772}. They embedded the Lasso in the Stability Selection framework and used the stability measure introduced by \citet{nogueira2018stability} to quantify the selection stability. They defined $\lambda_{\text{stable}}$ as the smallest $\lambda$ achieving a stability value of at least 0.75, and showed its Pareto optimality under two assumptions on the behaviour of the stability and accuracy curves across the regularization set. If $\lambda_{\text{stable}}$  is not attainable, they proposed selecting the smallest $\lambda$ whose corresponding stability remains within one standard deviation of the maximum observed stability, denoted by $\lambda_{\text{stable-1sd}}$. Since the goal of this paper is to obtain stable results, we adopt their approach for tuning $\lambda$ by jointly considering selection stability and prediction accuracy, rather than relying solely on cross-validation methods that prioritize prediction accuracy. Nonetheless, cross-validation-based methods remain applicable.

\section{Numerical Results}\label{s3}

We evaluate our method using several synthetic datasets and a real bioinformatics dataset. We used the corresponding GitHub repositories of \citet{AirHOLP}, \citet{nogueira2018stability}, and \citet{nouraie2025stability} to execute Air-HOLP, apply the stability measure, and perform variable decorrelation respectively. For regularization tuning, we used the GitHub repository associated with \citet{nouraie2024selection}. 

To create a set of candidate regularization values $\Lambda$, we use a $10$-fold cross-validation on the entire dataset $\mathcal{D}$. Although we use the sequence of tuning parameter values generated by 10-fold cross-validation in our numerical experiments, this choice is made purely for convenience. Any suitable sequence of candidate values, such as a logarithmically spaced grid up to $\lambda_{\max}$, may be used, since the proposed tuning procedure only requires evaluating the method over a sequence of candidate tuning parameters. The Air-HOLP method takes an input parameter, \texttt{Threshold}, which specifies the number of variables to retain during the screening process. In all experiments, this parameter was fixed at $n/\log(n)$ to eliminate its influence on the analysis. In addition, the maximum number of iterations for Air-HOLP was set to 10 across all experiments.

We use the \texttt{SGL} R package \citep{simon2018package} to fit the SGL and the \texttt{iilasso} R package\footnote{\url{https://github.com/tkdmah/iilasso/tree/master}} to fit the IILasso. For all Adaptive Lasso experiments, we fix the tuning parameter at $\gamma = 1$ to eliminate its influence on the analysis. In addition, to avoid division by zero when computing the adaptive weights, we add a small constant ($10^{-6}$) to the absolute values of the regression coefficients. In our implementation of Randomized Lasso, we set $\alpha = 0.2$ and $\Pr(w_j = \alpha) = 0.5$ for all $j$ values. It should be noted that, although Randomized Lasso is intended to assign random weights within Stability Selection \citep{meinshausen2010stability}, we observed that this implementation leads to very poor performance. Consequently, we opted to apply a fixed random weighting consistently across all sub-samples of the data.

\sloppy
We embed selection methods in the Stability Selection framework \citep{meinshausen2010stability} to evaluate their selection stability and accuracy. For methods employing variable-specific weights (penalty factors), the weights are obtained from the full dataset $\mathcal{D}$ and subsequently applied consistently across all models within the Stability Selection sub-samples. Simulation R scripts were executed using the \texttt{parallel} package to perform parallel computing across 7 CPU cores on a MacBook Air 13. 

\subsubsection*{Synthetic Data}
For the synthetic data experiments, we consider a challenging high-dimensional setting characterized by strong correlations among predictors. Specifically, datasets are generated with a sample size of $n = 100$ and $p = 1,000$ predictor variables. For each dataset, the predictor variables $\boldsymbol{x}^\top_{i}$ are independently drawn from the Normal distribution $\mathcal{N}(\boldsymbol{0}, \Sigma)$, where 
$
\Sigma \in \operatorname{mat}(p,p)$ has a diagonal of ones, and the off-diagonal elements are defined as follows
$$
\sigma_{jk} =
\begin{cases}  
\rho_g, & \text{if } j, k \text{ belong to the same group } G_g, \text{ where } g = 1, \dots, 5,  \\  
0, & \text{otherwise;}  
\end{cases}  
$$
where each group consists of consecutive indices
$$
G_g = \left\{ \frac{(g-1)p}{5} + 1, \dots, \frac{gp}{5} \right\}, \quad g = 1, \dots, 5.
$$
Each group $G_g$ is associated with a distinct covariance parameter $\rho_g$. Specifically, the covariance values are set as $\rho_1 = 0.8$, $\rho_2 = 0.85$, $\rho_3 = 0.9$, $\rho_4 = 0.95$, and $\rho_5 = 0.99$. Within each group, the highest-indexed variable is designated as relevant. The true regression coefficients of relevant variables are defined as $
\boldsymbol{\beta} = (3, 2.5, 2, 1.5, 1)^{\top}
$ following the order of the groups and there are $199$ irrelevant variables in each group. The error term $\boldsymbol{\varepsilon}$ is an i.i.d. sample from the standard Normal distribution $\mathcal{N}(0,1)$.

We generate 100 datasets with varying random seeds, following the setting described above. For each dataset, we generated $B=100$ sub-samples by randomly selecting $50\%$ of the observations without replacement, following the standard implementation of Stability Selection. In our implementation, we use $\lambda_{\text{stable-1sd}}$ for each dataset, as $\lambda_{\text{stable}}$ may not exist for some datasets. This approach ensures consistent regularization tuning across all datasets.

\begin{figure}[H]
    \centering
    \subfloat[Selection stability \label{fig1:fig1}]{
        \includegraphics[width=0.45\textwidth]{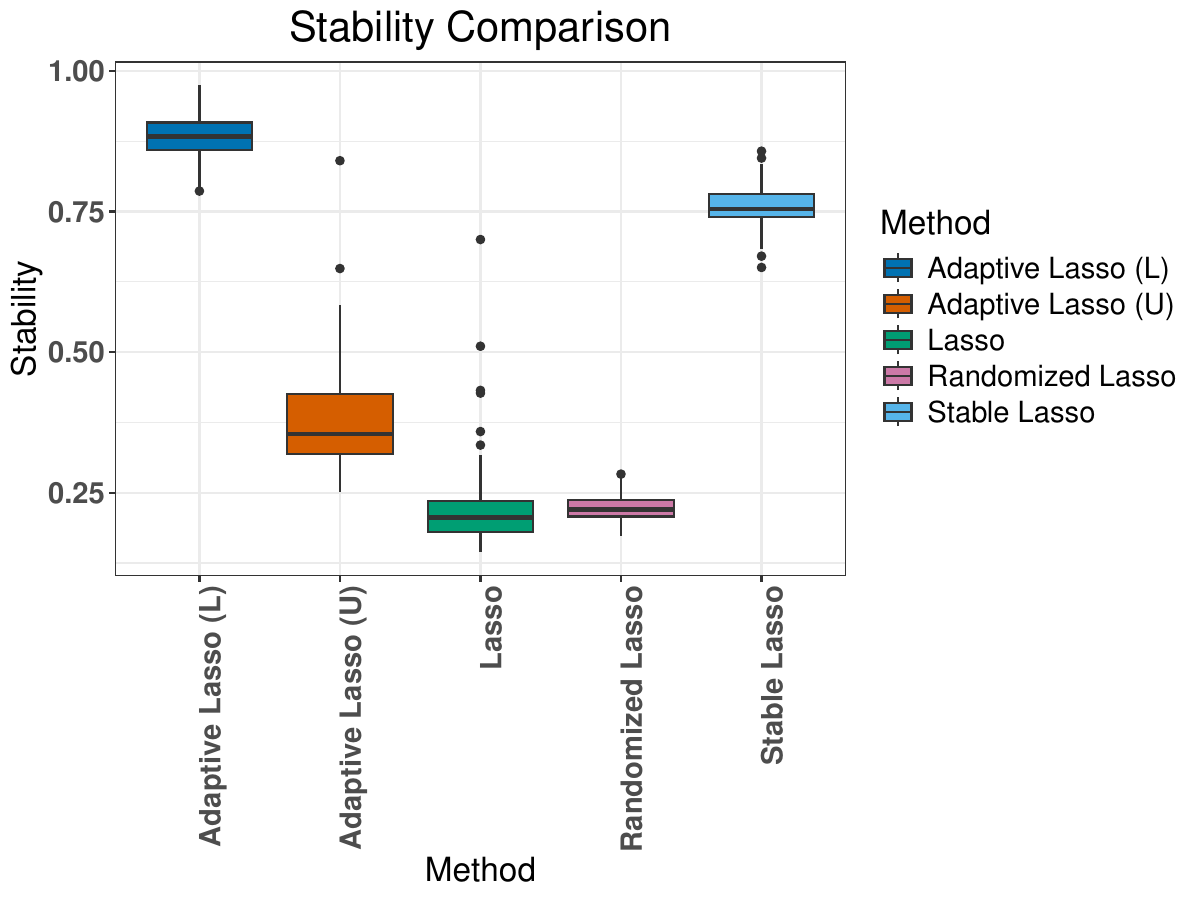}
    }
    \hfill
    \subfloat[Selection accuracy \label{fig1:fig2}]{
        \includegraphics[width=0.45\textwidth]{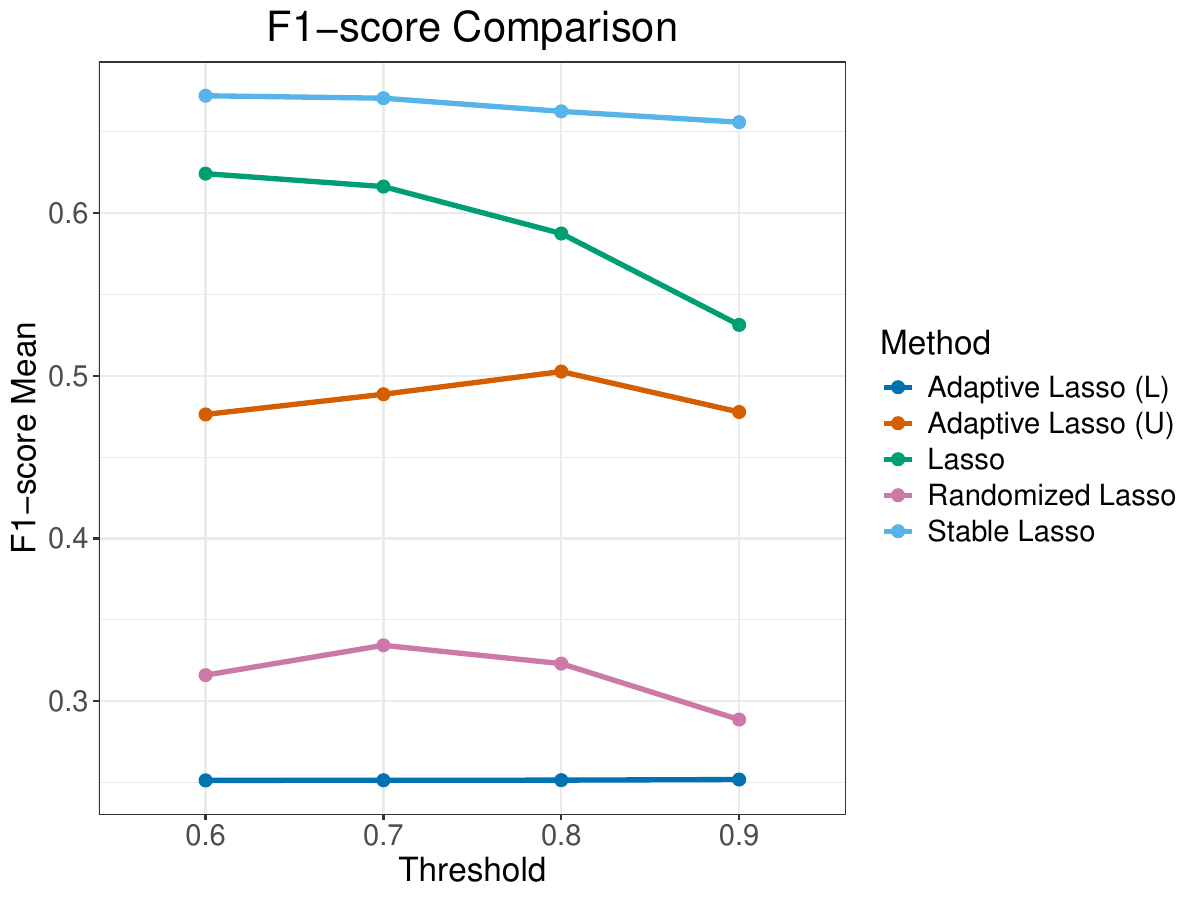}
    }
    \caption{Comparing Stability and F1-score between Stability Selection with Stable Lasso, Lasso, Adaptive Lasso, and Randomized Lasso}
    \label{fig1}
\end{figure}

Figure \ref{fig1} presents a comparative analysis of Stability Selection using different selection methods, evaluating both selection stability and average F1-score. The F1-score is defined as the harmonic mean of precision and recall, given by
\[
F1=\frac{2 \times\text{Precision}\times\text{Recall}}
{\text{Precision}+\text{Recall}},
\]
where
\[
\text{Precision}=\frac{TP}{TP+FP},
\qquad
\text{Recall}=\frac{TP}{TP+FN},
\]
with $TP$, $FP$, and $FN$ denoting the numbers of true positives, false positives, and false negatives, respectively. Here, the F1-score is computed to assess selection accuracy, treating the selection task as a binary classification problem. Figure \ref{fig1} compares the performance of Lasso, Stable Lasso, Adaptive Lasso with Lasso initial values (Adaptive Lasso (L)), Adaptive Lasso with univariate regression initial values (Adaptive Lasso (U)), and Randomized Lasso. 

Figure \ref{fig1:fig1} demonstrates that the Adaptive Lasso (L) achieves the highest stability among all methods, with Stable Lasso ranking second, attaining a median stability value of 0.75, indicative of excellent selection stability. In contrast, the Lasso exhibits the lowest median stability.
Figure \ref{fig1:fig2} presents the selection F1-scores of the methods across different Stability Selection thresholds $\pi_{\text{thr}}$. Stable Lasso consistently outperforms all alternatives across all threshold values. Notably, the Lasso ranks second, achieving higher accuracy than the Adaptive Lasso methods. In contrast, the Adaptive Lasso (L) demonstrates very poor selection accuracy despite its excellent stability.

\begin{figure}[H]
    \centering
    \subfloat[Stability - SGL \label{fig2:fig1}]{
        \includegraphics[width=0.45\textwidth]{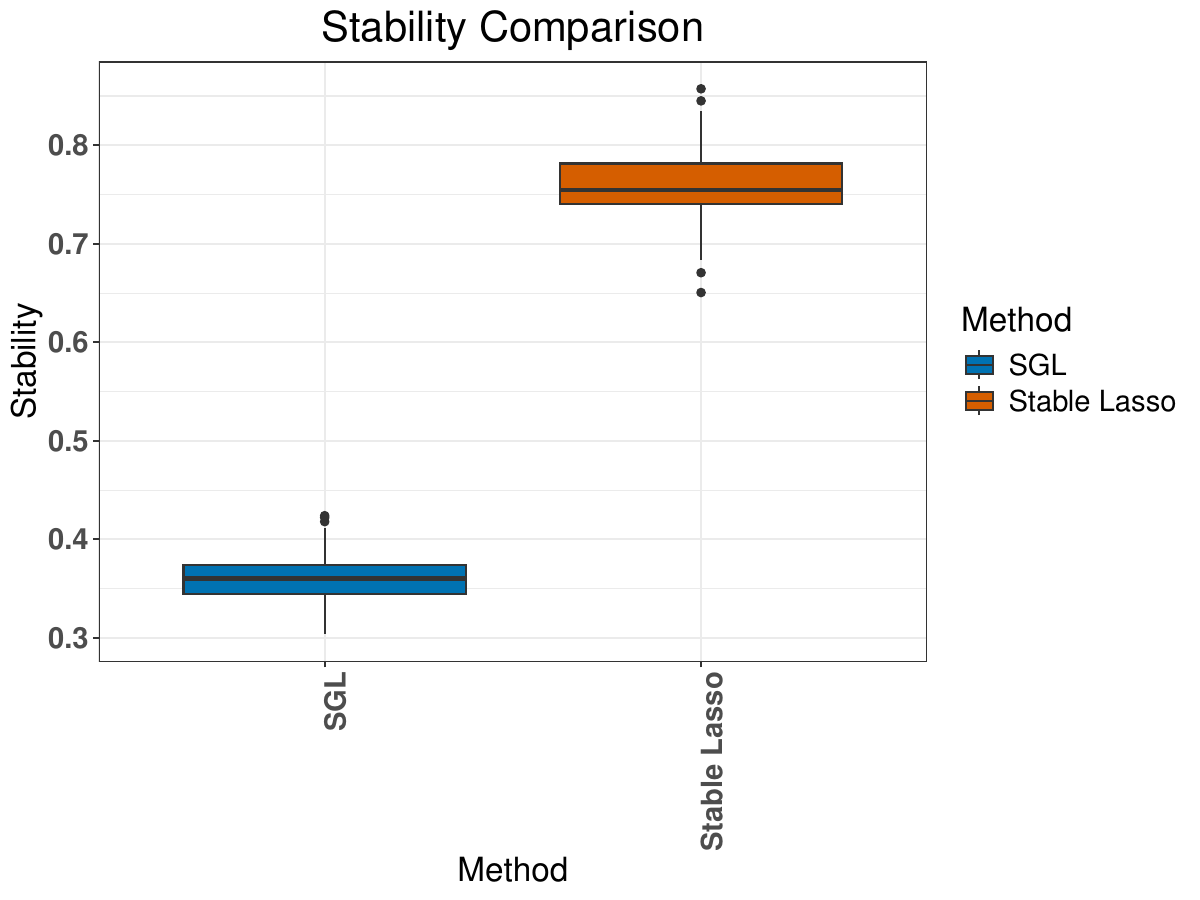}
    }
    \hfill
    \subfloat[Accuracy - SGL \label{fig2:fig2}]{
        \includegraphics[width=0.45\textwidth]{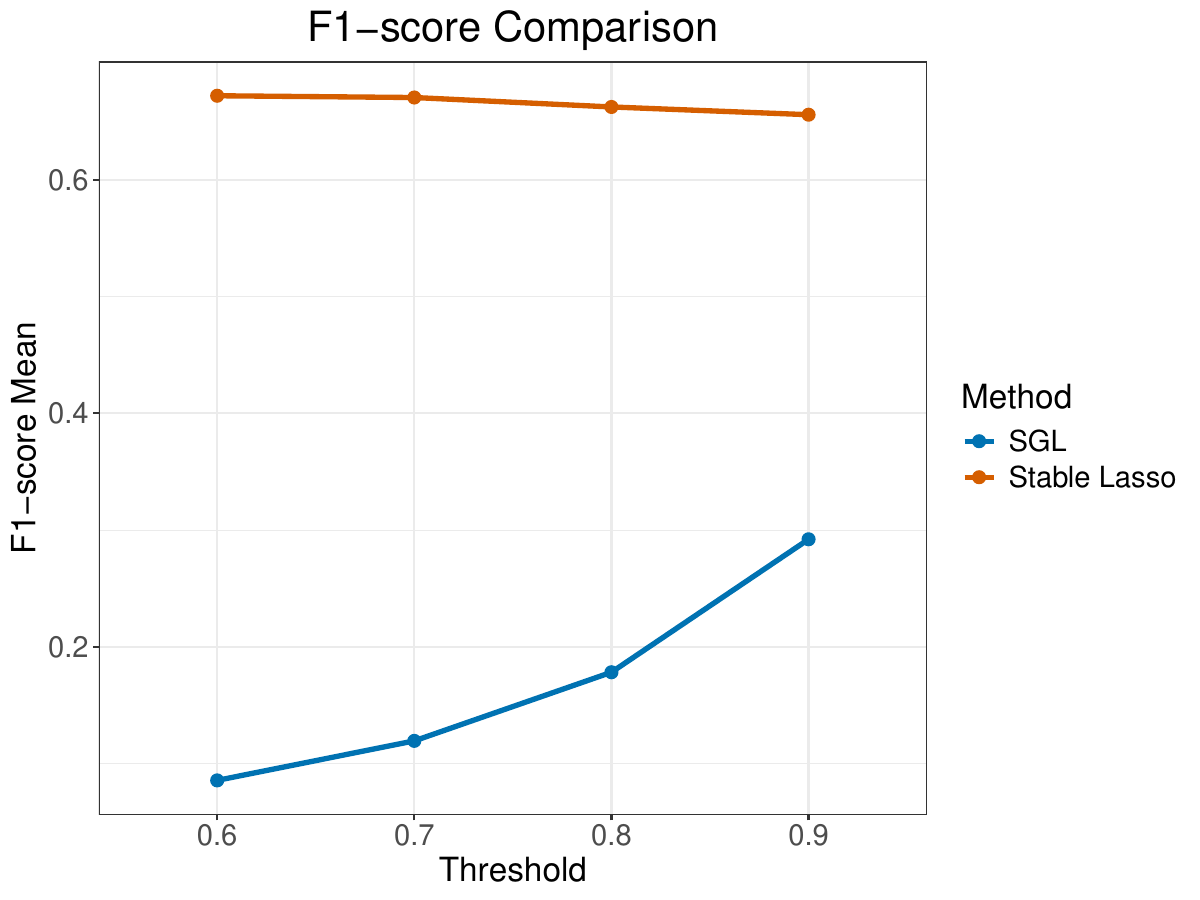}
    }

    \subfloat[Stability - IILasso \label{fig2:fig3}]{
        \includegraphics[width=0.45\textwidth]{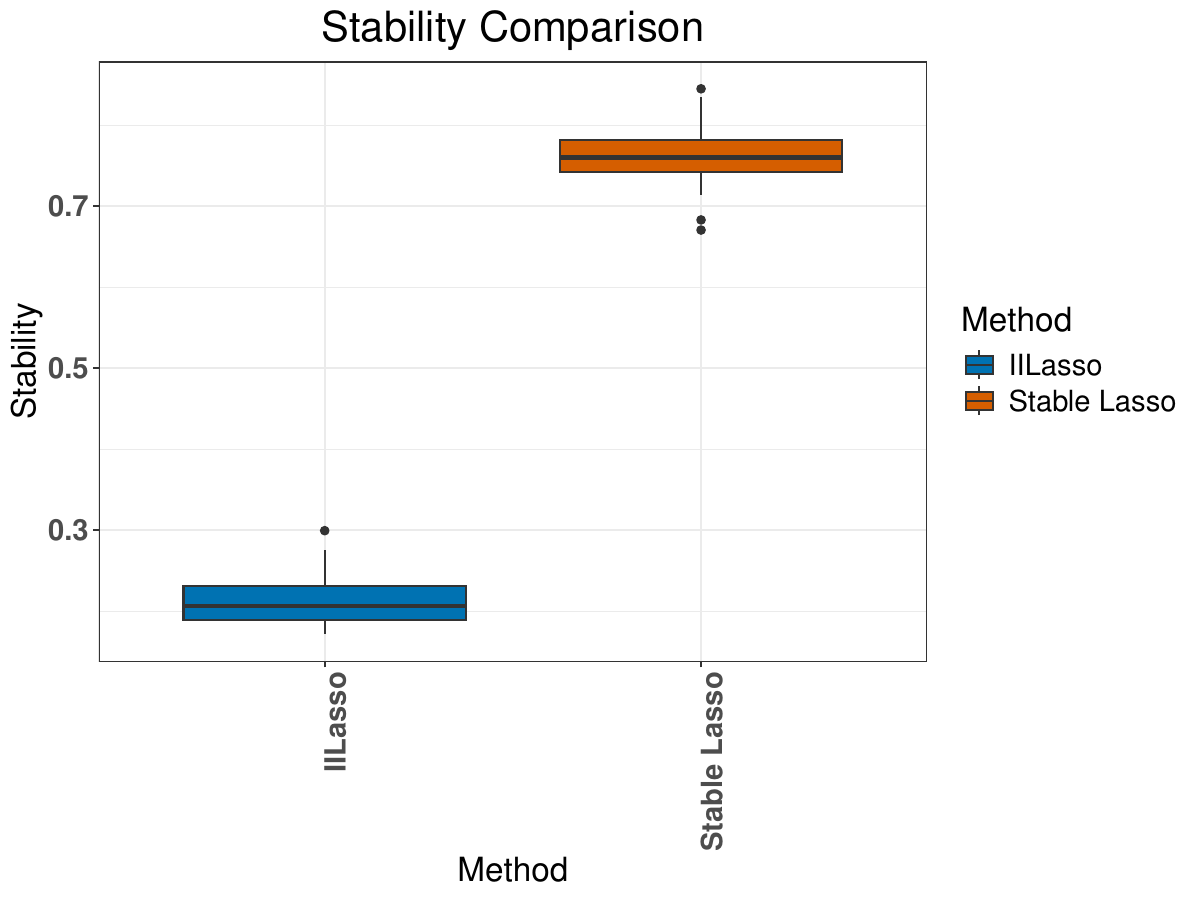}
    }
    \hfill
    \subfloat[Accuracy - IILasso \label{fig2:fig4}]{
        \includegraphics[width=0.45\textwidth]{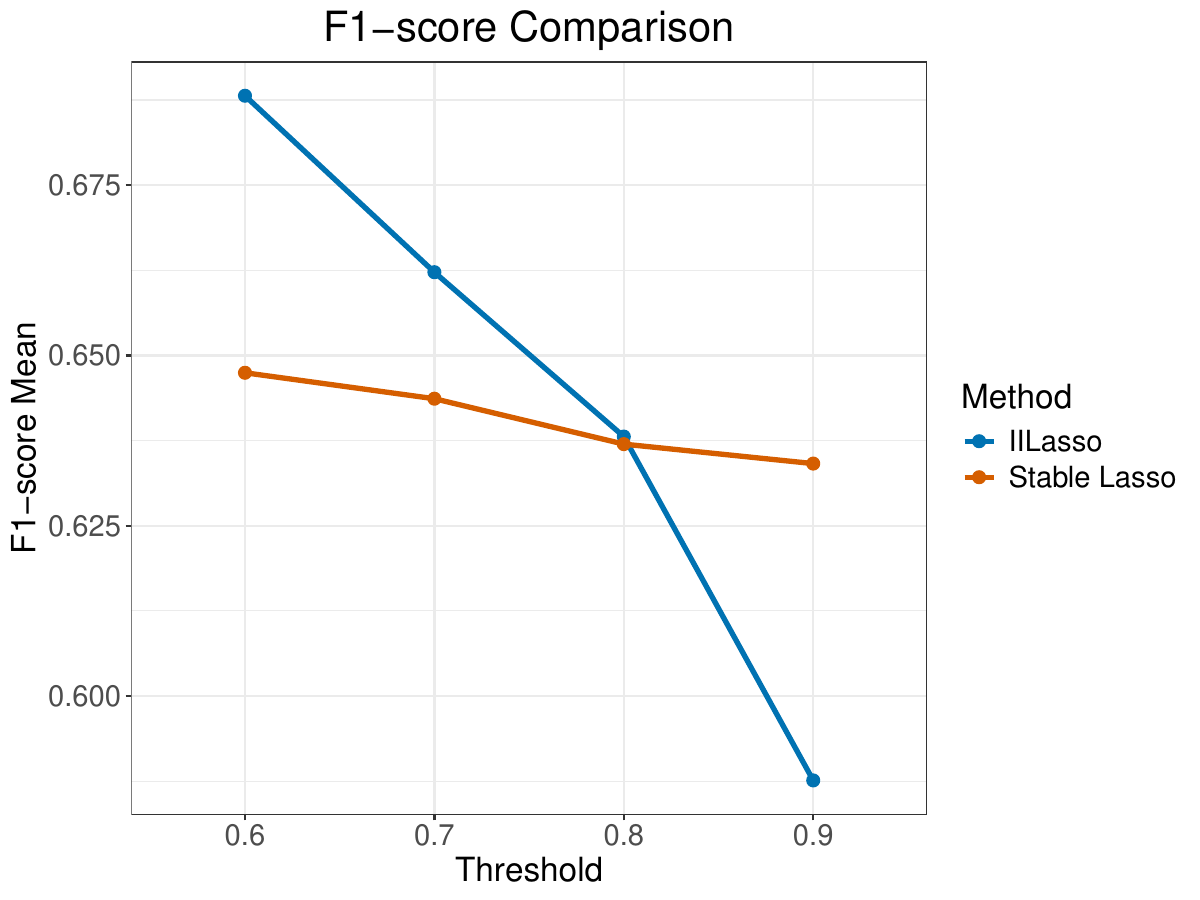}
    }

    \subfloat[Stability - Variable Decorrelation \label{fig2:fig5}]{
        \includegraphics[width=0.45\textwidth]{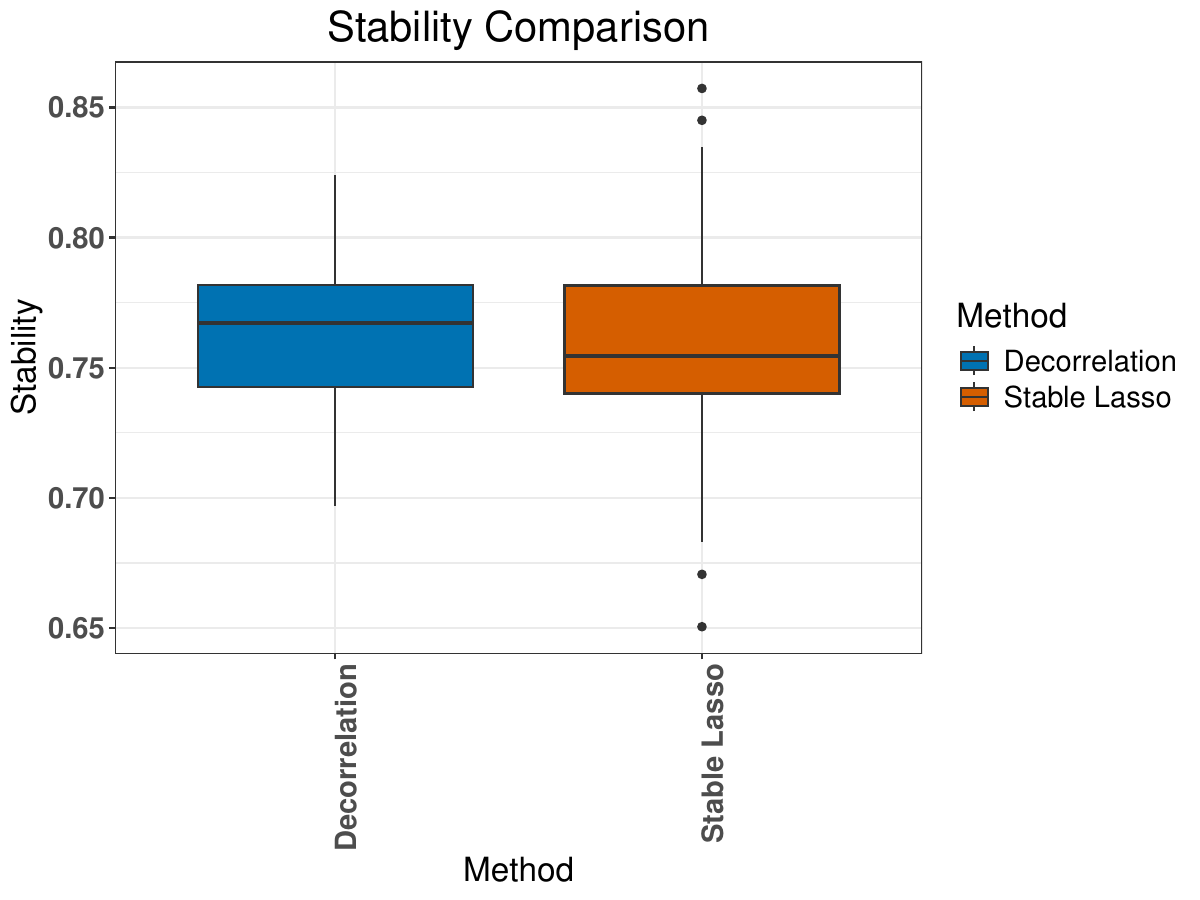}
    }
    \hfill
    \subfloat[Accuracy - Variable Decorrelation \label{fig2:fig6}]{
        \includegraphics[width=0.45\textwidth]{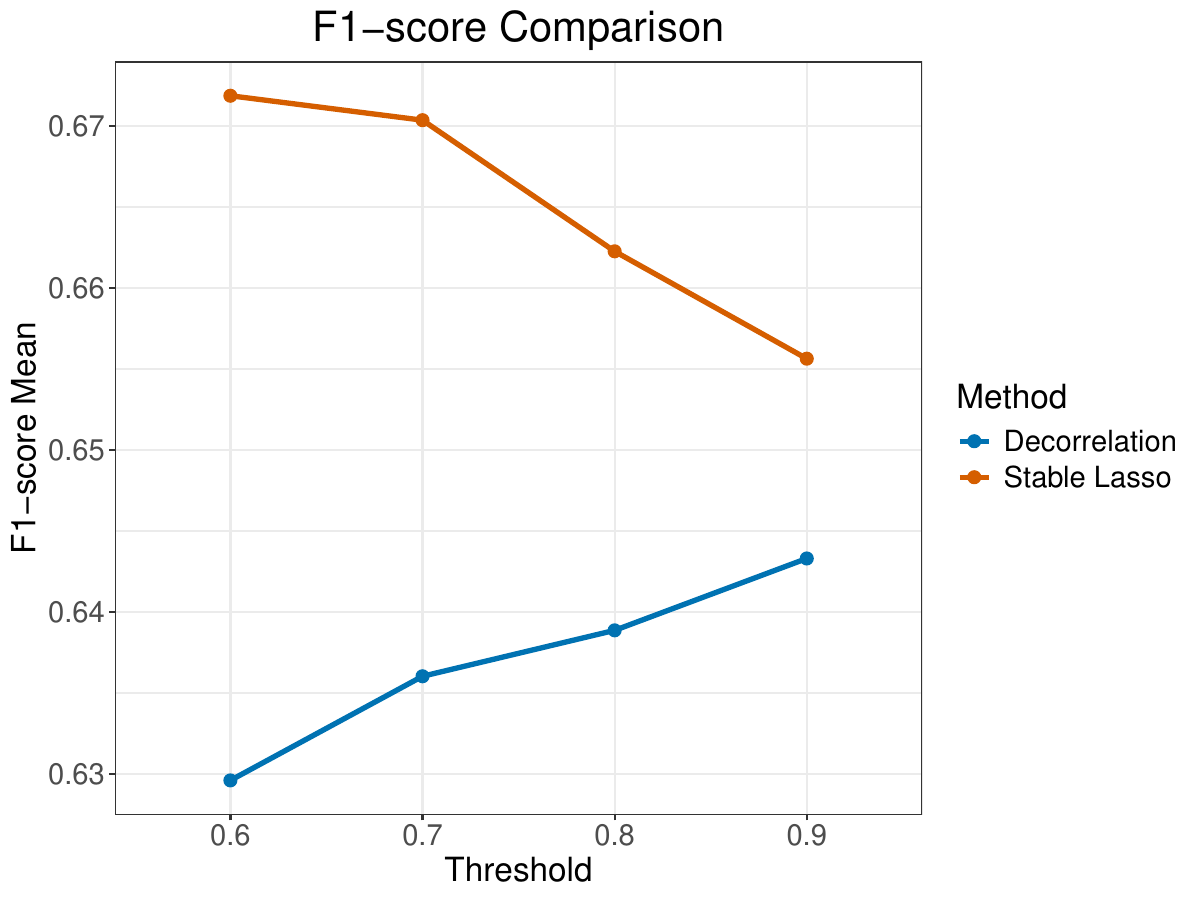}
    }

    \caption{Comparing Stability and F1-score between Stability Selection with Stable Lasso, SGL, IILasso, and variable decorrelation approach}
    \label{fig2}
\end{figure}

Figures \ref{fig2:fig1} and \ref{fig2:fig2} compare Stable Lasso with SGL, which requires prior knowledge of the group structure between variables. In our experiments, this group information was provided to SGL. It is important to note that SGL incurs a higher running time due to its complex optimization procedure. As shown in Figures \ref{fig2:fig1} and \ref{fig2:fig2}, Stable Lasso demonstrates significantly greater stability and accuracy than SGL. This highlights the effectiveness of our approach, even when compared to more complex models that leverage grouping structures among predictors.

Figures \ref{fig2:fig3} and \ref{fig2:fig4} compare Stable Lasso with IILasso, which exploits the full covariance structure of the predictors. Due to the high computational cost of IILasso's complex optimization, results are based on 25 simulated datasets, in contrast to the 100 datasets used in the other figures. Notably, even after reducing the number of datasets and employing parallel computing, this experiment still required just over 72 hours to complete.  As shown in Figure \ref{fig2:fig3}, Stable Lasso exhibits substantially higher stability than IILasso. Figure \ref{fig2:fig4} shows that the selection accuracy of the two models is comparable, with IILasso performing  better at lower thresholds and Stable Lasso better at higher thresholds. Moreover, the selection accuracy of Stable Lasso is more robust to the choice of threshold compared to IILasso. These results highlight the effectiveness of our computationally efficient approach, even when compared to sophisticated methods that exploit the full covariance structure among predictors.

Figures \ref{fig2:fig5} and \ref{fig2:fig6} compare Stable Lasso with the variable decorrelation approach proposed by \citet{nouraie2025stability}. As shown in Figure \ref{fig2:fig5}, both methods achieve excellent stability, with the variable decorrelation approach showing slightly higher median stability values. This is particularly noteworthy, as their method applies Lasso to an orthonormal surrogate matrix, whereas Stable Lasso attains comparable stability while operating directly on the original, highly correlated data. Figure \ref{fig2:fig6} indicates that their selection accuracy is also comparable, with Stable Lasso outperforming the alternative across all threshold values.

An additional advantage of Stable Lasso over the variable decorrelation method is interpretability. While the selected components in the decorrelation approach serve as surrogates for the original variables, Stable Lasso operates directly in the original data space, selecting the relevant predictors. These results demonstrate that Stable Lasso can deliver stable and accurate variable selection in complex and challenging settings, outperforming several sophisticated alternatives.

\begin{figure}[H]
    \centering
    \subfloat[Stability - Low dimension \label{fig3:fig1}]{
        \includegraphics[width=0.45\textwidth]{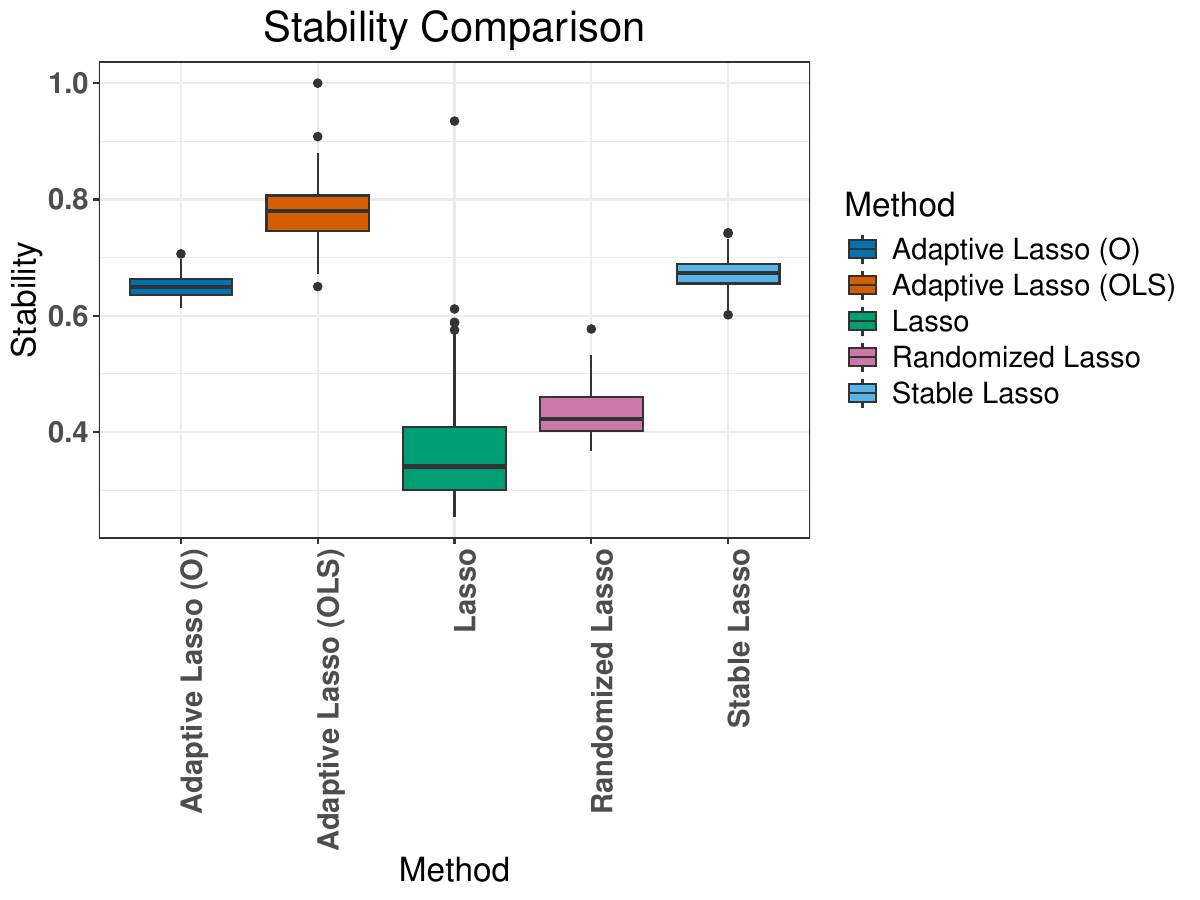}
    }
    \hfill
    \subfloat[Accuracy - Low dimension \label{fig3:fig2}]{
        \includegraphics[width=0.45\textwidth]{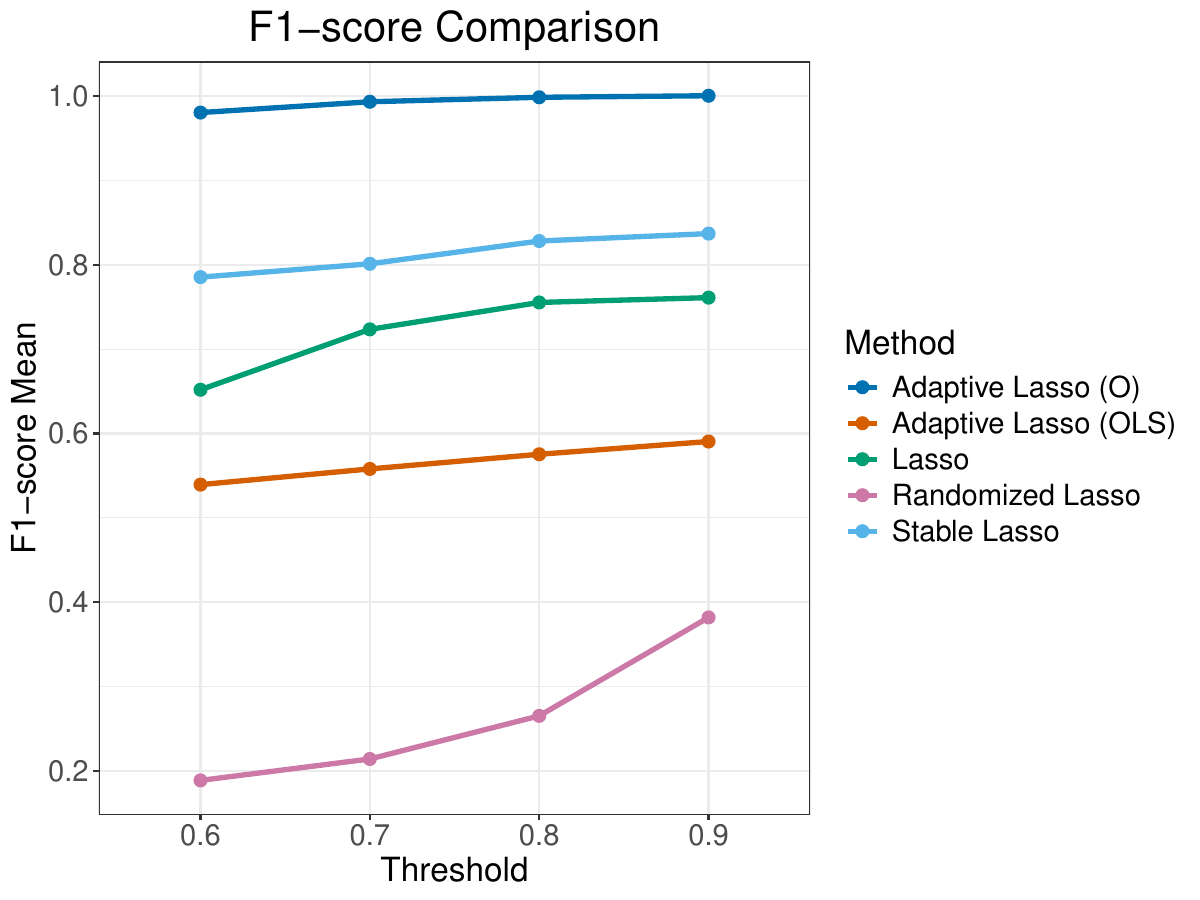}
    }
    \\
    \subfloat[Stability - Low correlation \label{fig3:fig3}]{
        \includegraphics[width=0.45\textwidth]{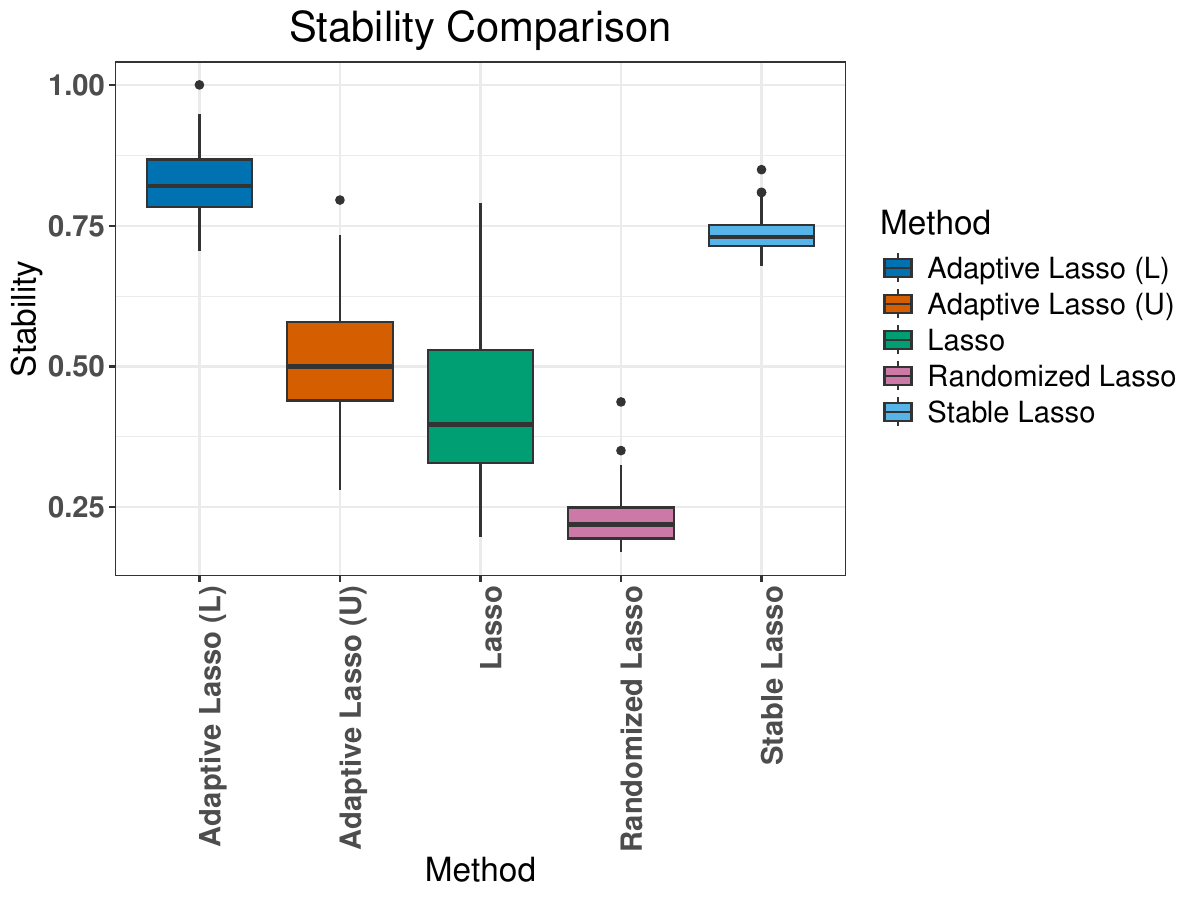}
    }
    \hfill
    \subfloat[Accuracy - Low correlation \label{fig3:fig4}]{
        \includegraphics[width=0.45\textwidth]{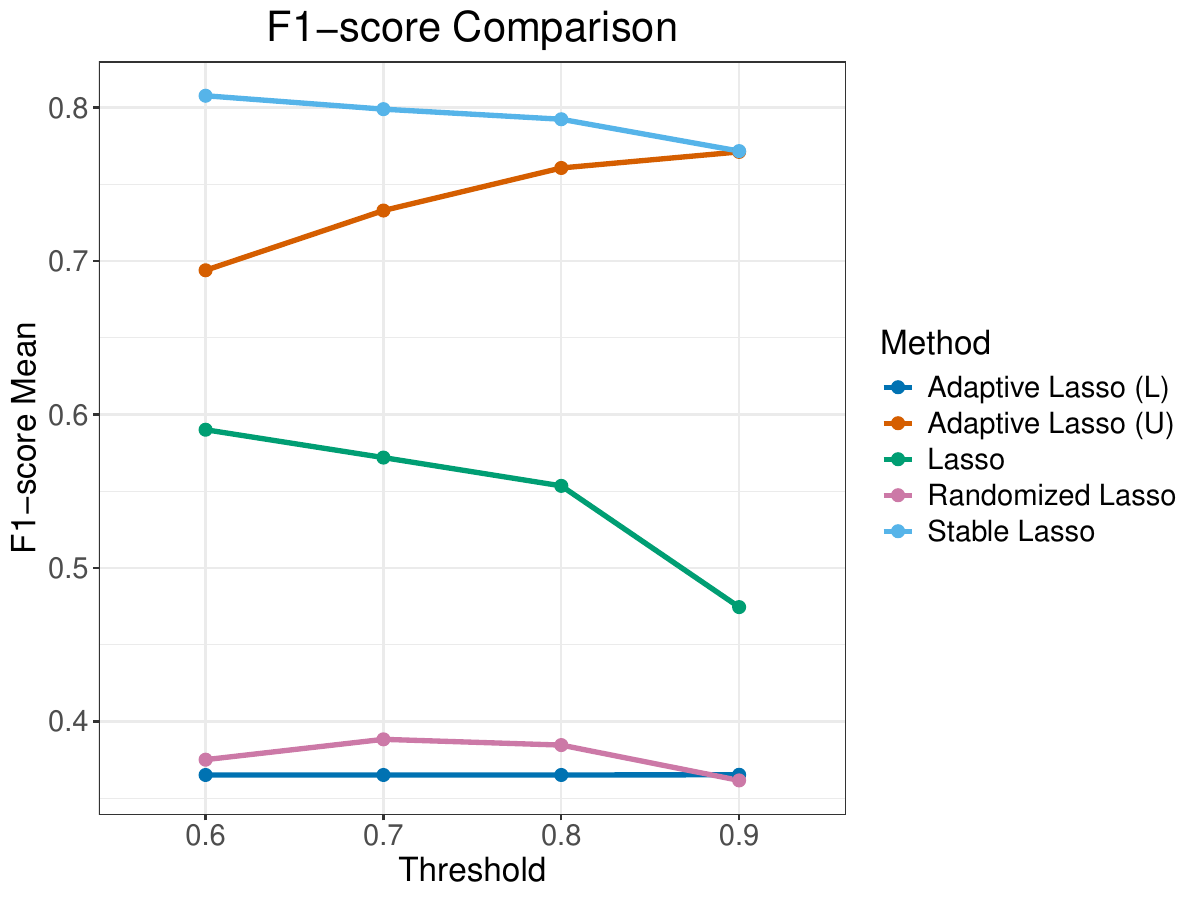}
    }
    \caption{Comparing Stability and F1-score between Stability Selection with Stable Lasso,
Lasso, Adaptive Lasso, and Randomized Lasso in the low-dimension, and low-correlation scenarios}
    \label{fig5}
\end{figure}

Now we extend the initial data generation setting to explore a reduced-dimensional setting, decreasing the number of variables from $1,000$ to $80$ while preserving the original group structure. This modification allows us to assess the performance of our method in a low-dimensional context. This setting is of particular interest because it falls within the low-dimensional setting, where Ordinary Least Squares (OLS) estimation of regression coefficients is feasible. In the figures, the adaptive Lasso method powered by OLS estimates is labeled Adaptive Lasso (OLS). In addition, we consider an oracle version of the Adaptive Lasso, denoted by Adaptive Lasso (O), which employs the true regression coefficients used to generate the data. The results for this scenario are presented in Figures~\ref{fig3:fig1} and~\ref{fig3:fig2}. 

Figure~\ref{fig3:fig1} shows that Adaptive Lasso (OLS) achieves the highest stability, with Stable Lasso ranking second, slightly above Adaptive Lasso (O).  Figure~\ref{fig3:fig2} further demonstrates that Adaptive Lasso (O), benefiting from oracle information, attains the highest selection accuracy, followed by Stable Lasso. These results underscore the effectiveness of our method in the low-dimensional setting.

In addition, we consider a scenario with attenuated covariance values to examine our method's performance under less extreme correlation structures. This scenario retains the original high-dimensional setting with $n = 100$ and $p = 1,000$, but reduces the correlation levels among the predictors. This analysis aims to determine whether the strong results reported earlier are specific to extremely correlated data, or if they remain robust under more moderate correlation structures. Specifically, we set the correlation parameters to $\rho_1 = 0.4$, $\rho_2 = 0.5$, $\rho_3 = 0.6$, $\rho_4 = 0.7$, and $\rho_5 = 0.8$. The results for this scenario are presented in Figures~\ref{fig3:fig3} and~\ref{fig3:fig4}.

As shown in Figure~\ref{fig3:fig3}, Adaptive Lasso (L) exhibits the highest stability, followed by Stable Lasso. Figure~\ref{fig3:fig4} shows that Stable Lasso achieves the highest average F1-score across all decision thresholds, followed by Adaptive Lasso (U). These results indicate that the performance advantages of Stable Lasso persist even under moderate correlation structures.

\subsubsection*{Comparison of Initial Ranking Methods}

We compare the performance of Stable Lasso using three initial ranking methods: HOLP, Ridge-HOLP, and Air-HOLP under the original high-dimensional, highly correlated data-generating scenario. For HOLP, a small ridge regularization term ($r = 10^{-6}$) is used in place of zero to avoid numerical instability caused by computational singularity.

\begin{figure}[H]
    \centering
    \subfloat[Selection stability \label{fig:comp1}]{
        \includegraphics[width=0.45\textwidth]{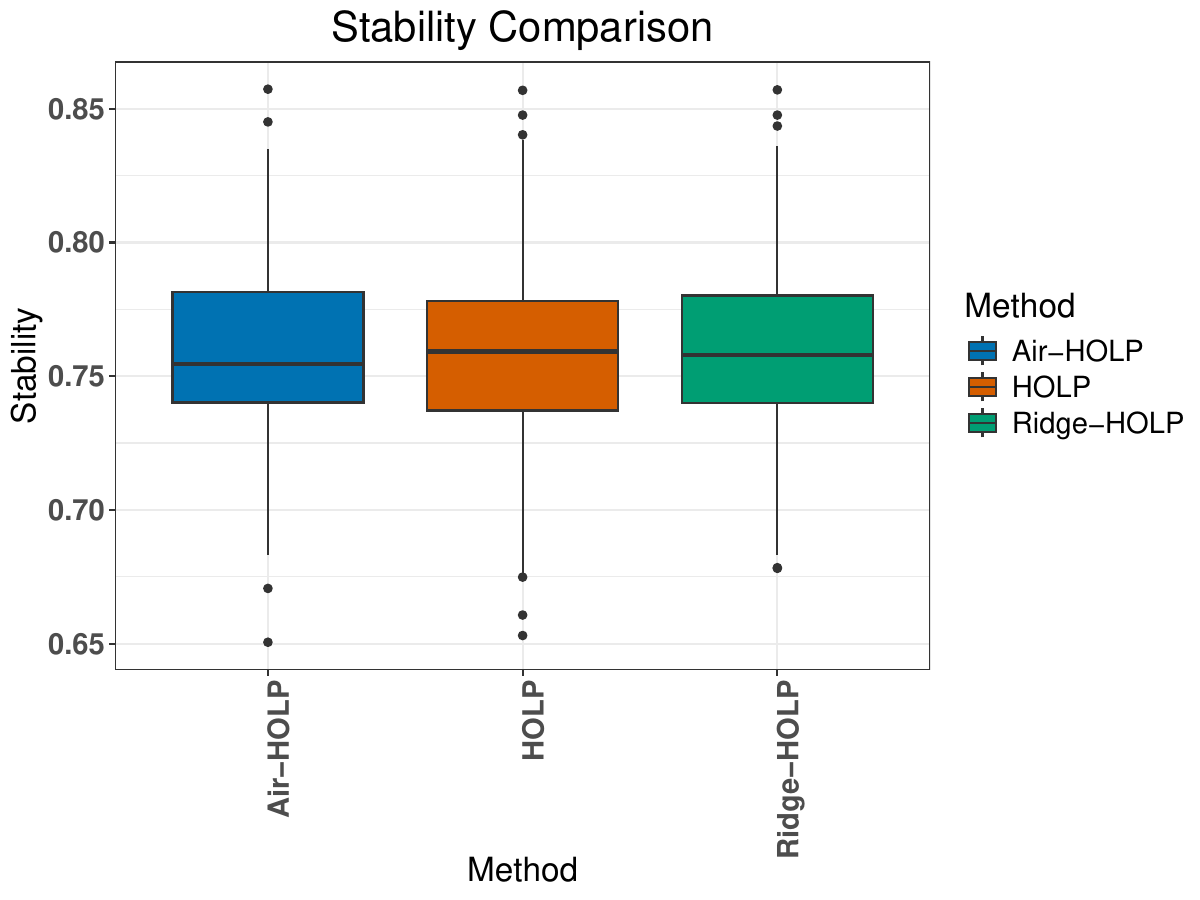}
    }
    \hfill
    \subfloat[Selection accuracy \label{fig:comp2}]{
        \includegraphics[width=0.45\textwidth]{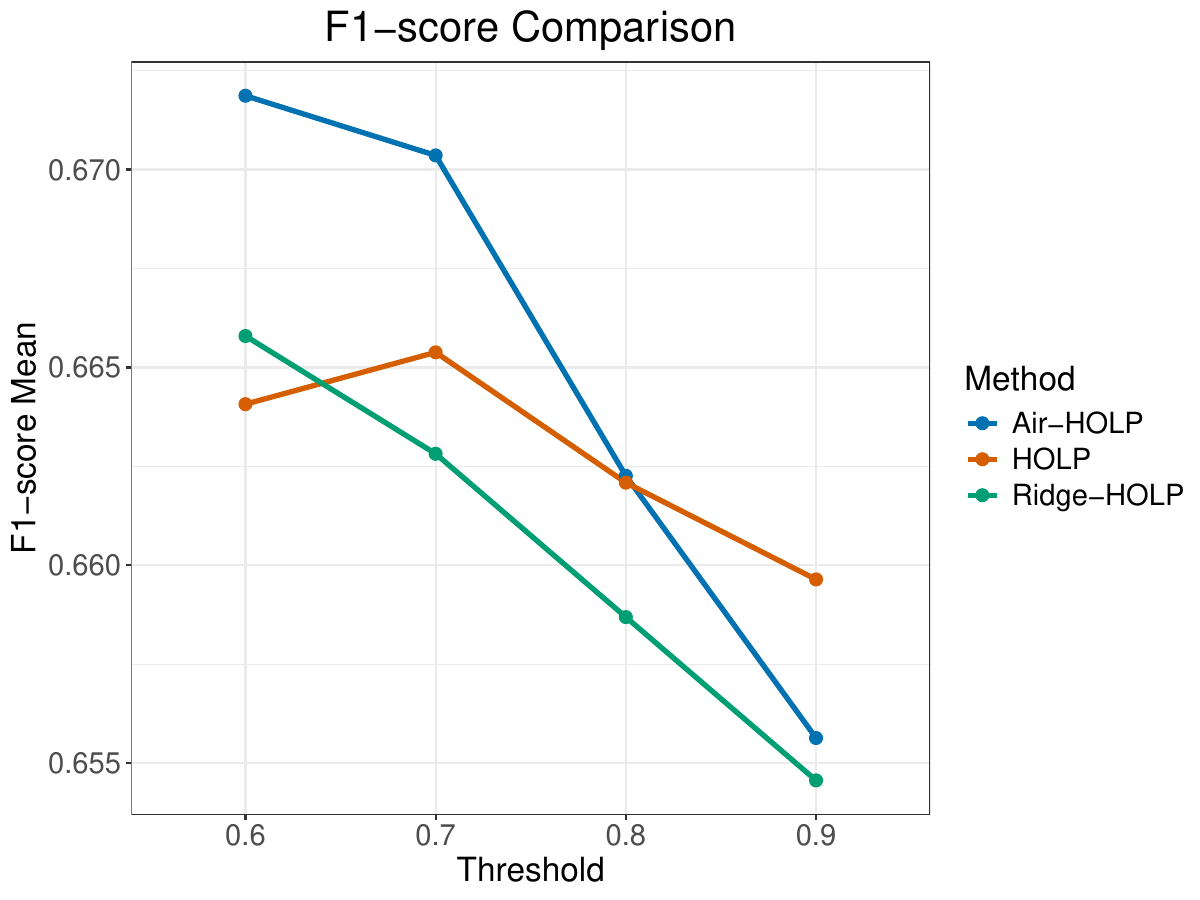}
    }
    \caption{Comparison of Stability and F1-score for Stability Selection with Stable Lasso using three different initial ranking methods}
    \label{fig_comp}
\end{figure}

As shown in Figure~\ref{fig:comp1}, all three methods produce nearly identical stability values. This is consistent with Remark~\ref{remark1}, which explains that stability is primarily determined by the weighting scheme rather than the initial ranking method. While the stability levels are similar, differences may still arise in variable selection accuracy. Figure~\ref{fig:comp2} shows that the three methods achieve comparable selection performance, with Air-HOLP consistently performing slightly better than Ridge-HOLP across all thresholds and marginally outperforming HOLP at lower thresholds, in line with its emphasis on predictive performance.

Overall, Figure~\ref{fig_comp} suggests that, in this simulation setting, the choice of initial ranking method has only a modest effect on the performance of Stable Lasso. Although the three approaches differ algorithmically, they exhibit broadly similar behavior within our framework. The observed performance differences are relatively small and may depend on the particular simulation scenario. In general, Ridge-HOLP may offer improved numerical stability when $XX^\top$ is close to degeneracy or when $p$ is close to $n$ \citep{10.1111/rssb.12127}.

\subsubsection*{Comparison of $c$ Values}

Section~\ref{s2} introduced the one-parameter family of weighting functions
\[
w(r_{j})=1-r_{j}^{-c},
\]
where the parameter $c>0$ controls the rate at which the weights approach the standard Lasso penalty as the rank increases. Throughout the experiments we set $c=1$ for simplicity and interpretability. To assess the sensitivity of the proposed method to this modeling choice, we repeat the original simulation study for
$c \in \{0.25,\,0.5,\,1,\,2,\,4\}$.

\begin{figure}[H]
    \centering
    \subfloat[Selection stability \label{fig:comp3}]{
        \includegraphics[width=0.45\textwidth]{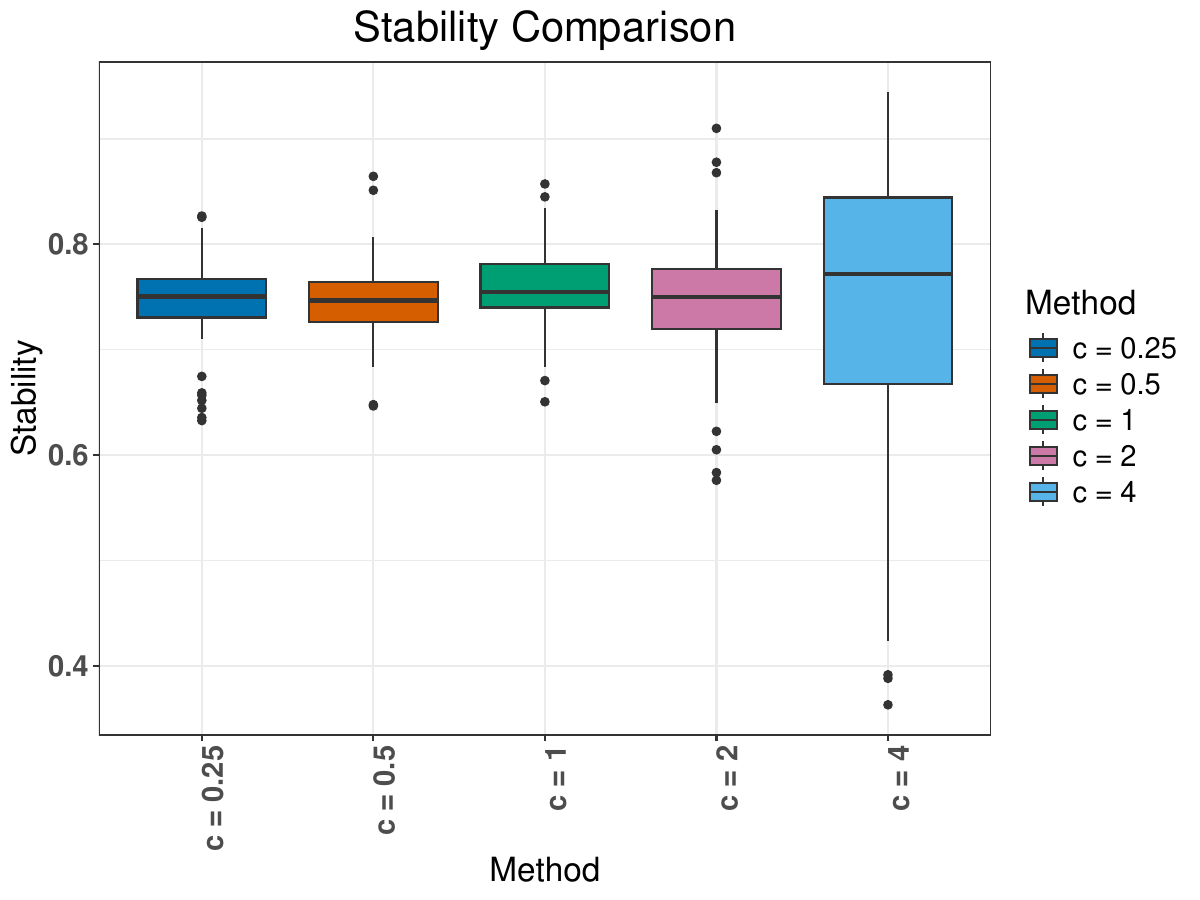}
    }
    \hfill
    \subfloat[Selection accuracy \label{fig:comp4}]{
        \includegraphics[width=0.45\textwidth]{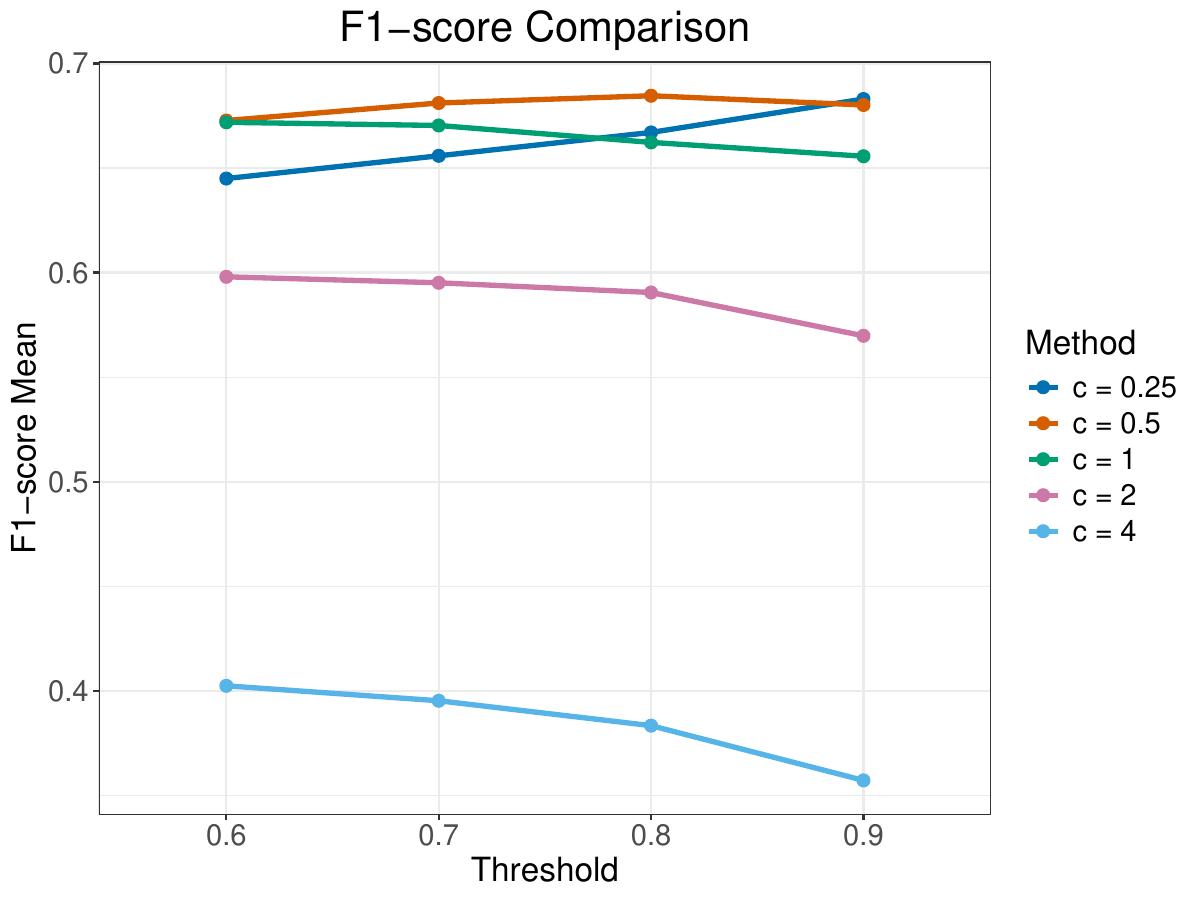}
    }
    \caption{Comparison of Stability and F1-score for Stability Selection with Stable Lasso for different values of $c$}
    \label{fig_compc}
\end{figure}

Figure~\ref{fig:comp3} compares the stability values obtained for the different choices of $c$. Overall, the method appears relatively insensitive to the choice of $c$, although larger values tend to produce greater variability in the stability measure across simulation replicates.

Figure~\ref{fig:comp4} compares the variable selection performance for the same values of $c$. The results indicate that $c\leq1$ yields broadly comparable performance, whereas larger values ($c>1$) tend to reduce selection accuracy. Among the values considered, $c=0.5$ achieves the best overall performance across the selection thresholds, although the improvement over $c=1$ is modest. These results indicate that the proposed method is relatively robust to the choice of $c$ over a broad range of values. In practice, a grid of candidate values for $c$ may be considered, with the final choice determined according to the resulting variable selection performance.

\subsubsection*{Comparison with Other Monotone Functional Forms}

In Section~\ref{s2}, we adopted the constant-elasticity specification as a parsimonious modeling assumption for constructing the weighting function. Here, we compare this proposal with three alternative monotone weighting schemes. For our proposed family, we fix the decay parameter at $c=1$, yielding $w(r_j)=1-r_j^{-1}$. The competing alternatives are
\[
w(r_j)=1-\frac{1}{1+\log(r_j)}, \qquad
w(r_j)=1-\frac{1}{1+r_j}, \qquad
w(r_j)=1-e^{-r_j}.
\]
Among these functions, the exponential form converges most rapidly to one and therefore quickly approaches the unweighted Lasso, whereas the logarithmic form converges the most slowly. The remaining reciprocal form is a slight modification of our proposal that assigns marginally larger penalties to higher-ranked variables.

\begin{figure}[H]
    \centering
    \subfloat[Selection stability \label{fig:comp5}]{
        \includegraphics[width=0.45\textwidth]{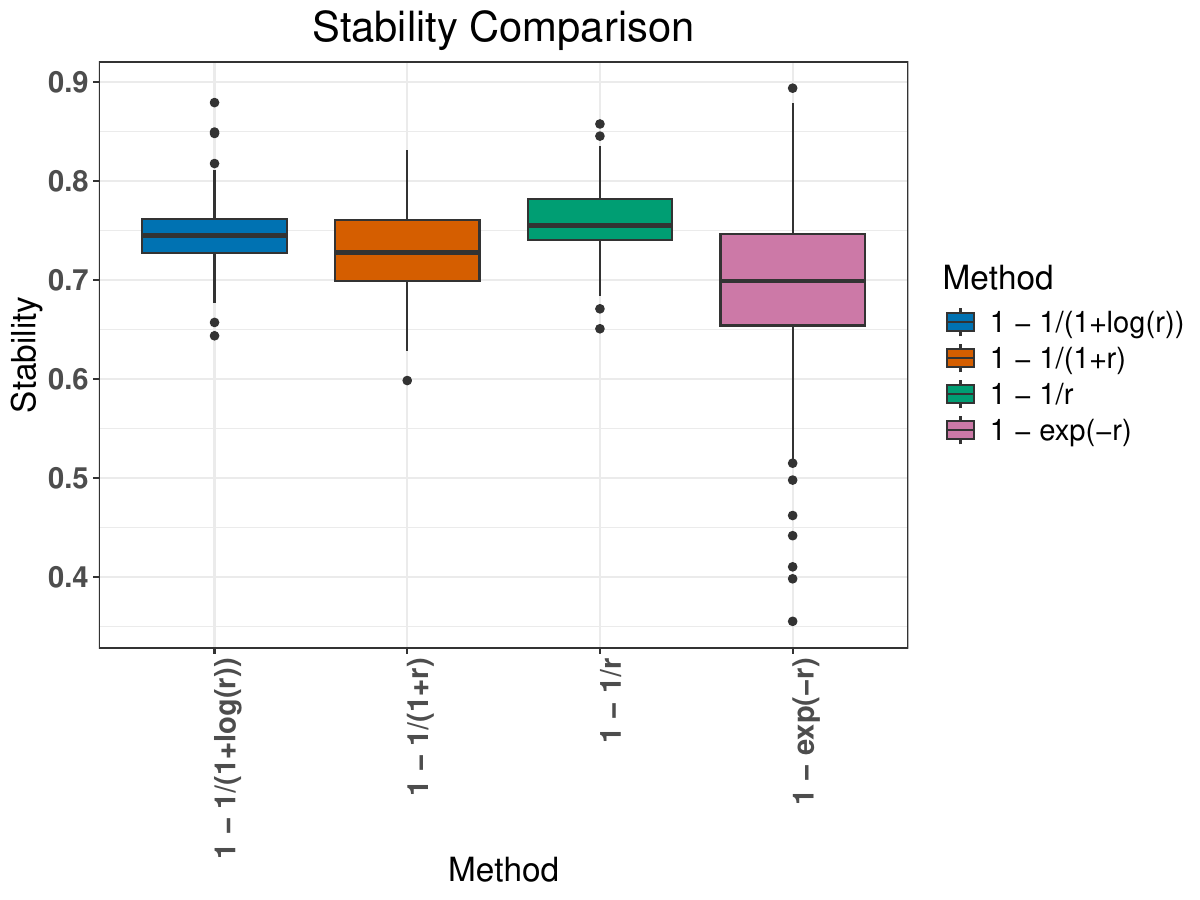}
    }
    \hfill
    \subfloat[Selection accuracy \label{fig:comp6}]{
        \includegraphics[width=0.45\textwidth]{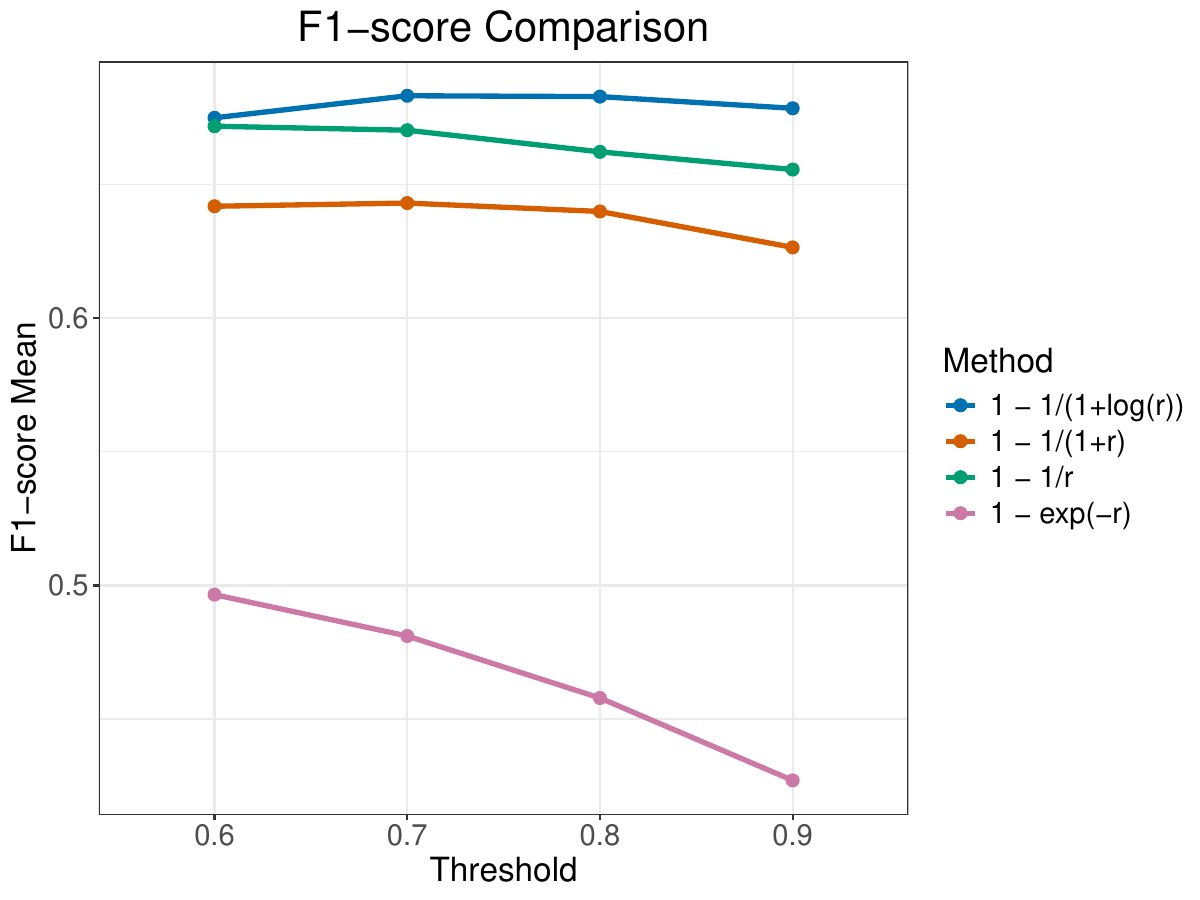}
    }
    \caption{Comparison of Stability and F1-score for Stability Selection with Stable Lasso with different weighting functions $w(r_j)$}
    \label{fig_compc_func}
\end{figure}

Figure~\ref{fig:comp5} shows that all four weighting schemes achieve broadly comparable stability. Nevertheless, our proposed weighting exhibits the highest median stability. As expected, the exponential weighting, which is closest to the unweighted Lasso due to its rapid convergence to one, produces both lower stability and greater variability. Figure~\ref{fig:comp6} indicates that the logarithmic weighting and our proposal provide the strongest variable selection performance. In this simulation, the logarithmic weighting performs slightly better, consistent with the results in Figure~\ref{fig:comp4}, where smaller values of the decay parameter (e.g., $c=0.5$) yielded improved performance. This suggests that, for the present data-generating scenario, a slower decay of the weighting function may be beneficial. Notably, such behavior can be achieved within the proposed constant-elasticity framework simply by selecting a smaller value of $c$.

\subsubsection*{Mice PCR Data}

As a real-world example, we employ a Polymerase Chain Reaction (PCR) dataset, which serves to validate and verify the effectiveness of our proposed approach. \citet{10.1371/journal.pgen.0020006} conducted
an experiment which examines the genetics of two inbred mouse populations (C57BL/6J and
BTBR). A total of $n = 60$ second generation (F2) samples, with 31 female and 29 male mice, were used to monitor
the expression levels of $p = 22,575$ genes. Some physiological phenotypes, including numbers of
phosphoenopyruvate carboxykinase, glycerol-3-phosphate acyltransferase and stearoyl-CoA
desaturase 1 were measured by quantitative realtime PCR. In this example,
we study the relationship between phosphoenopyruvate carboxykinase as the response variable and the
gene expression levels. The gene expression data is standardized and the response vector is centered before the analysis. \citet{10.1111/rssb.12095} approached this problem with a split-and-merge Bayesian approach and identified \texttt{1429089\_s\_at},
\texttt{1430779\_at}, \texttt{1432745\_at}, \texttt{1437871\_at}, \texttt{1440699\_at}, and \texttt{1459563\_x\_at} as relevant genes. Later \citet{10.1214/22-BA1351} identified \texttt{1438937\_x\_at} and \texttt{1438936\_s\_at} as relevant genes. 

We set the number of Stability Selection sub-samples $B = 100$. Applying Lasso within Stability Selection on this dataset produced a maximum stability value of $0.054$ (rounded to three decimals), which is very low. This indicates that using Stability Selection alone, even with careful regularization tuning, does not yield stable results, highlighting the need for a more advanced selection approach. Here, with $\lambda_{\text{stable-1sd}} = 0.394$ and $\pi_{\text{thr}} = 0.2$, the gene $\texttt{1438937\_x\_at}$ is identified as relevant, with a selection frequency of 0.35 under $\lambda_{\text{stable-1sd}}$. This gene was also identified by \citet{10.1214/22-BA1351}.

Applying Stable Lasso resulted in a maximum selection stability of $0.85$ (rounded to three decimals), substantially higher than the previous approach. In this case, $\lambda_{\text{stable}}$ exists and is $0.433$. Using this $\lambda_{\text{stable}}$, the genes $\texttt{1437871\_at}$ and $\texttt{1438937\_x\_at}$ were identified with selection frequencies of 1 and 0.88, respectively. The first gene was reported by \citet{10.1111/rssb.12095}, while the latter was identified by \citet{10.1214/22-BA1351}.

\section{Discussion and Conclusion}\label{s4}
In this section, we briefly discuss some related methods which, while not explicitly framed within the context of stability, fall within the broader scope of this paper. Our approach shares certain similarities with these methods but differs fundamentally in how stability is defined and quantified, thereby underscoring the unique contribution of our method.

Sorted L-One Penalized Estimation \citep[SLOPE;][]{bogdan2015slope} introduces a ranking-based penalty on the regression coefficients, encouraging variable selection by ordering predictors according to their estimated effect magnitudes during the optimization process. While conceptually related to our approach, \citet{6840355} demonstrated that SLOPE generalizes the Octagonal Shrinkage and Clustering Algorithm for Regression \citep[OSCAR;][]{bondell2008simultaneous}, which is designed to perform grouping selection. This contrasts with the objective of the present work, which emphasizes exclusive selection. \citet{feser2023sparse} introduced Sparse Group SLOPE, which performs bi-level selection in a manner similar to the SGL. However, as with our discussion of SGL, we emphasize that our work differs from theirs.

\citet{chatterjee2025univariate} introduced Univariate-Guided Sparse Regression, a two-step procedure that preserves the signs of univariate regression coefficients throughout the optimization process and incorporates their magnitudes. Although there are some conceptual similarities between their approach and ours, the two methods are fundamentally different and address distinct objectives. While their method is designed to enhance sign stability of the Lasso, our focus is on improving its selection stability.  

Furthermore, our approach, which enhances the selection stability of the Lasso by incorporating prior information derived from a variable ranking method, bears conceptual resemblance to Bayesian variable selection methods, such as those proposed by \citet{park2008bayesian} and \citet{nouraie2026bayesian}.

\begin{figure}[H]
    \centering
    \subfloat[Stability - SCAD \label{fig4:fig1}]{
        \includegraphics[width=0.45\textwidth]{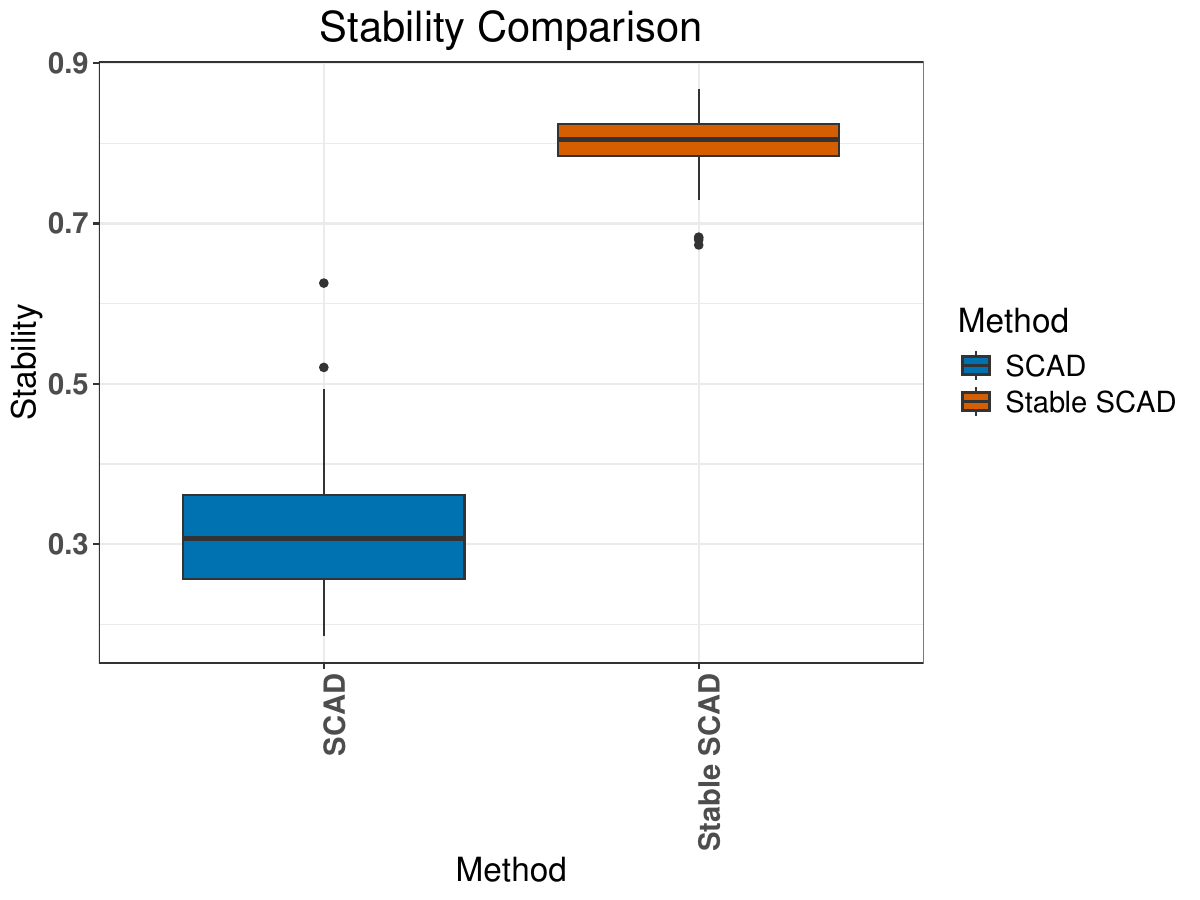}
    }
    \hfill
    \subfloat[Accuracy - SCAD \label{fig4:fig2}]{
        \includegraphics[width=0.45\textwidth]{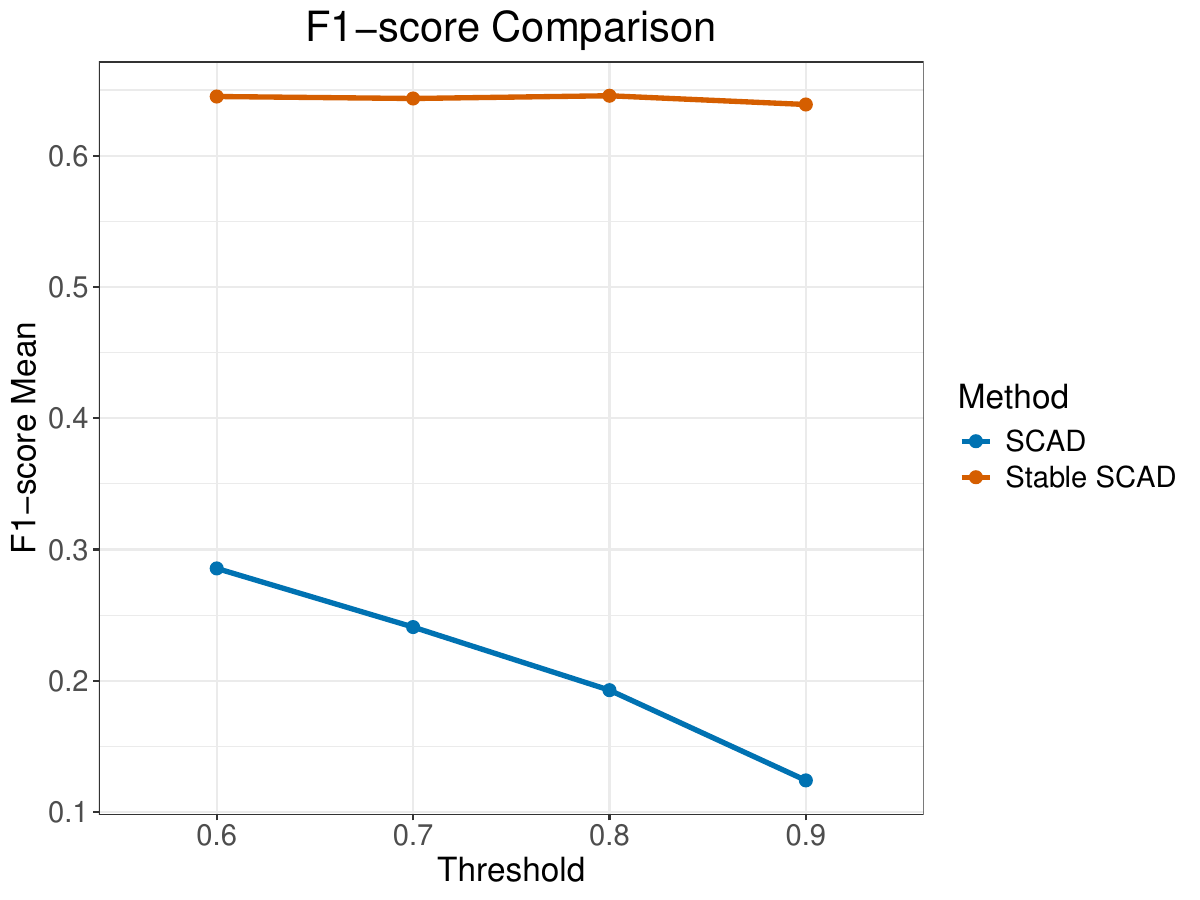}
    }
    \\
    \subfloat[Stability - MCP \label{fig4:fig3}]{
        \includegraphics[width=0.45\textwidth]{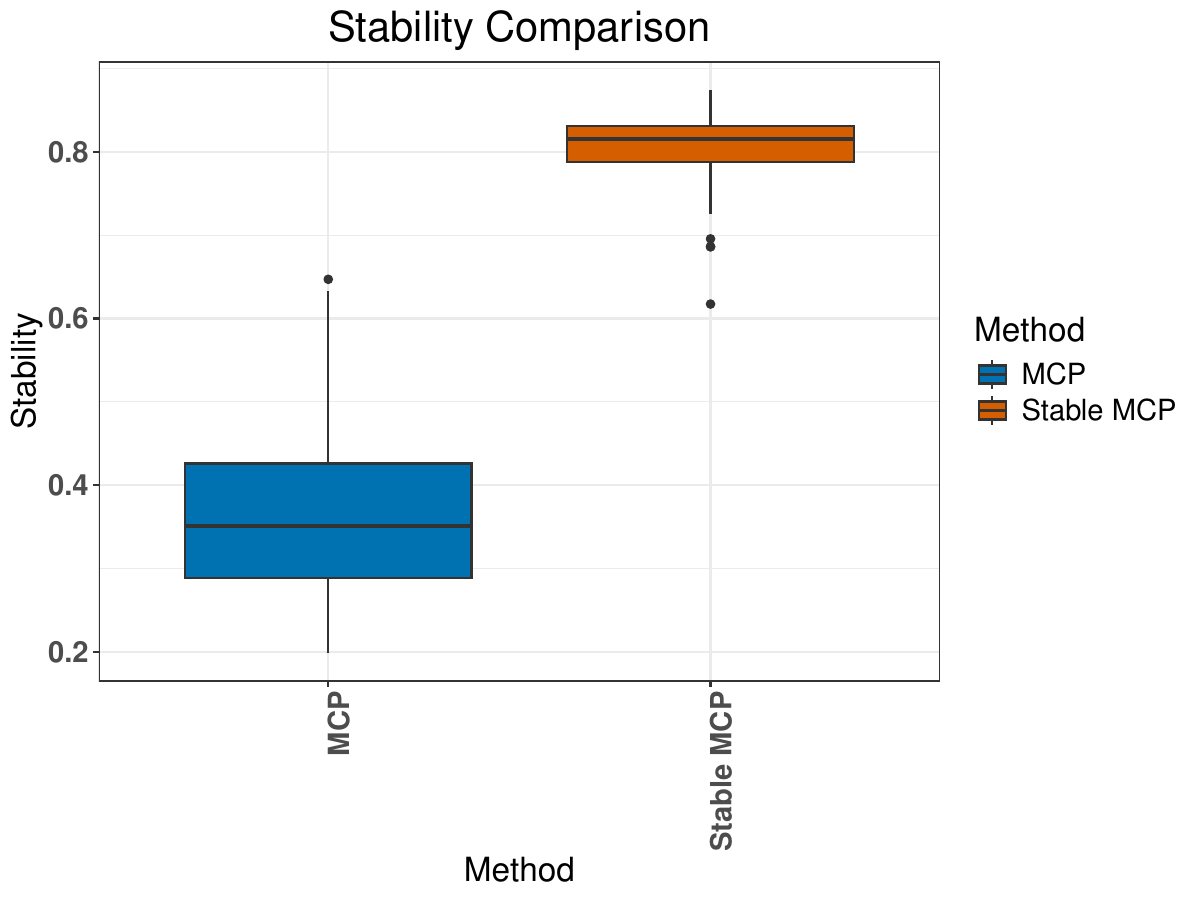}
    }
    \hfill
    \subfloat[Accuracy - MCP \label{fig4:fig4}]{
        \includegraphics[width=0.45\textwidth]{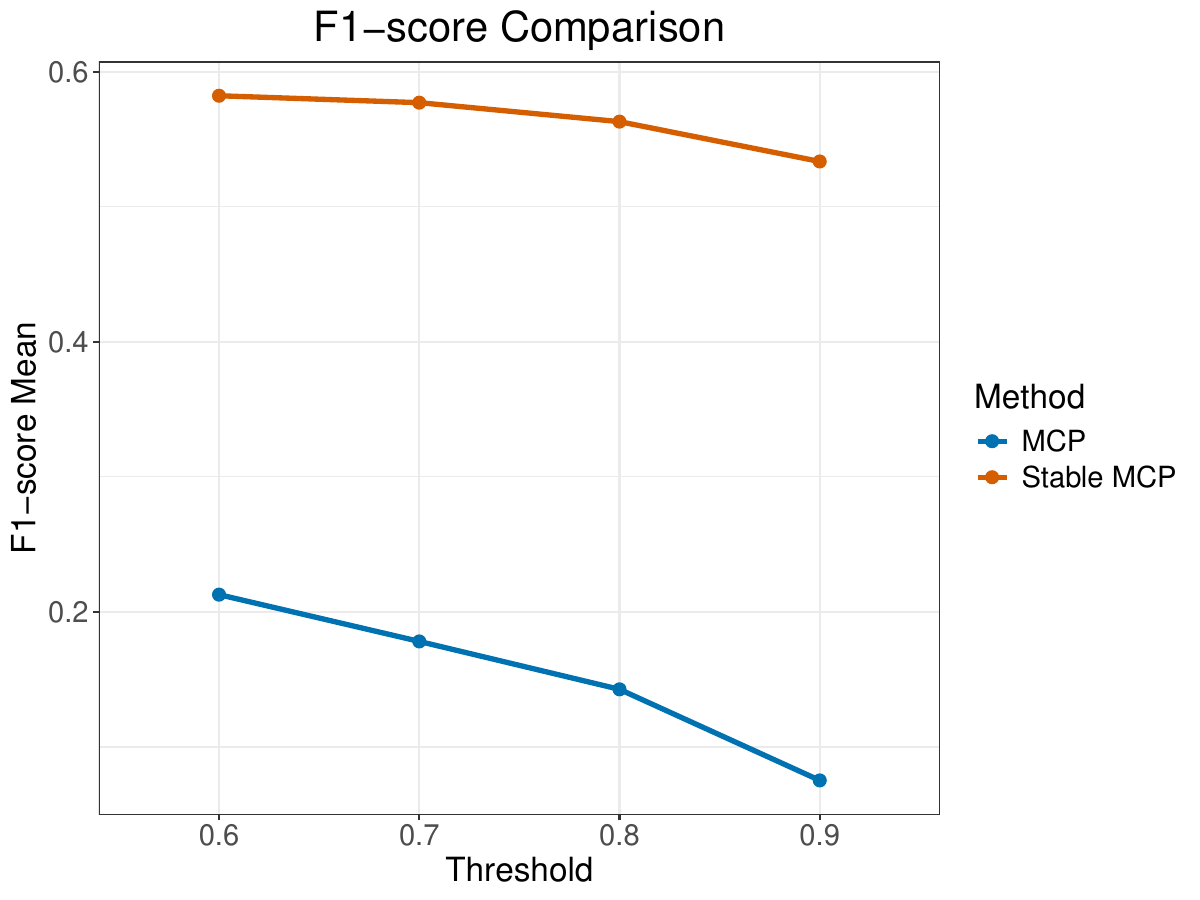}
    }
    \caption{Comparing Stability and F1-score between Stability Selection with Stable SCAD,
SCAD, Stable MCP, and MCP}
    \label{fig4}
\end{figure}

Finally, we present numerical results obtained by applying our proposed weighting method to the Smoothly Clipped Absolute Deviation (SCAD; \citealt{Fan01122001}) and Minimax Concave Penalty (MCP; \citealt{MCP}) methods. We adopt the primary simulation scenario with $n = 100$ and $p = 1,000$. The outcomes of this experiment are summarized in Figure~\ref{fig4}. Analogous to our implementation of the method as `Stable Lasso', we refer to SCAD and MCP when augmented with the weights defined in Equation~\eqref{eqn-weights} as `Stable SCAD' and `Stable MCP', respectively. Computations for SCAD and MCP were performed using the \texttt{ncvreg} R package \citep{ncvreg}. As illustrated in Figures~\ref{fig4:fig1} and \ref{fig4:fig2}, SCAD equipped with our proposed weighting scheme exhibits substantially improved stability and accuracy compared with the standard SCAD. Similarly, Figures~\ref{fig4:fig3} and \ref{fig4:fig4} demonstrate the same enhancement for MCP. These results underscore the potential of our approach as a general-purpose solution for stable variable selection, motivating further investigation and development.

\section*{Competing Interests}
The authors declare that they have no conflict of interest.

\section*{Author Contributions Statement}
 
Mahdi Nouraie was responsible for drafting the manuscript, the development of the research methodology and for writing the computer code used throughout. Samuel Muller and Houying Zhu provided critical feedback on the content of the manuscript, refining the clarity and scope of the manuscript and the computer code.  

\section*{Data and Code Availability}
The mice PCR dataset is accessible JRSS(B) Datasets Vol.77(5) at \url{https://fdslive.oup.com/www.oup.com/academic/content/rss-datasets/jrsssb/77-5-2015-B.zip}. 

The source code used for the paper is accessible through the following GitHub repository: \url{https://github.com/MahdiNouraie/Stable-Lasso}.

\section*{Acknowledgements}
We would like to thank the Associate Editor and the reviewers for their valuable comments and suggestions, which helped improve the quality and clarity of this manuscript.

\section*{Funding}
Mahdi Nouraie was supported by the Macquarie University Research Excellence Scholarship (20213605). Samuel Muller and Houying Zhu were supported by the Australian Research Council Discovery Project Grant (DP260100348).

\bibliographystyle{bst/sn-mathphys-num}  
\bibliography{sn-bibliography}

@article{tibshirani1996regression,
  title={Regression {S}hrinkage and {S}election via the {L}asso},
  author={Tibshirani, Robert},
  journal={Journal of the Royal Statistical Society Series B: Statistical Methodology},
  volume={58},
  number={1},
  pages={267--288},
  year={1996},
}

@article{AirHOLP,
    author = {Ibrahim Joudah and Samuel Muller and Houying Zhu},
    title = {{Air-HOLP: adaptive regularized feature screening for high dimensional
correlated data}},
    journal = {Statistics and Computing},
    volume = {35},
    number = {63},
    year = {2025}
}

@article{10.1111/rssb.12127,
    author = {Wang, Xiangyu and Leng, Chenlei},
    title = {{High Dimensional Ordinary Least Squares Projection for Screening Variables}},
    journal = {Journal of the Royal Statistical Society Series B: Statistical Methodology},
    volume = {78},
    number = {3},
    pages = {589-611},
    year = {2015}
}

@article{nogueira2018stability,
  title={On the {S}tability of {F}eature {S}election {A}lgorithms},
  author={Nogueira, Sarah and Sechidis, Konstantinos and Brown, Gavin},
  journal={Journal of Machine Learning Research},
  volume={18},
  number={174},
  pages={1--54},
  year={2018}
}

@article{nouraie2026bayesian,
  title={Bayesian stability selection and inference on selection probabilities},
  author={Nouraie, Mahdi and Smith, Connor and Muller, Samuel},
  journal={Statistics and Computing},
  volume={36},
  number={62},
  year={2026},
  publisher={Springer}
}

@article{wang2025ridge,
  title={Ridge partial correlation screening for ultrahigh-dimensional data},
  author={Wang, Run and Nguyen, An and Dutta, Somak and Roy, Vivekananda},
  journal={arXiv preprint arXiv:2504.19393},
  year={2025}
}

@article{meinshausen2006high,
  title={{High-dimensional graphs and variable selection with the Lasso}},
  author={Meinshausen, Nicolai and B{\"u}hlmann, Peter},
  journal = {The Annals of Statistics},
  volume = {34},
  number = {3},
  pages = {1436--1462},
  year={2006}
}

@article{zou2005regularization,
  title={Regularization and variable selection via the elastic net},
  author={Zou, Hui and Hastie, Trevor},
  journal={Journal of the Royal Statistical Society Series B: Statistical Methodology},
  volume={67},
  number={2},
  pages={301--320},
  year={2005},
}

@article{shah2013variable,
  title={Variable selection with error control: another look at stability selection},
  author={Shah, Rajen D and Samworth, Richard J},
  journal={Journal of the Royal Statistical Society Series B: Statistical Methodology},
  volume={75},
  number={1},
  pages={55--80},
  year={2013},
  publisher={Oxford University Press}
}

@article{bogdan2015slope,
  title={{SLOPE—adaptive variable selection via convex optimization}},
  author={Bogdan, Ma{\l}gorzata and Van Den Berg, Ewout and Sabatti, Chiara and Su, Weijie and Cand{\`e}s, Emmanuel J},
  journal={The Annals of Applied Statistics},
  volume={9},
  number={3},
  pages={1103--1140},
  year={2015}
}

@article{tibshirani2005sparsity,
  title={{Sparsity and smoothness via the fused Lasso}},
  author={Tibshirani, Robert and Saunders, Michael and Rosset, Saharon and Zhu, Ji and Knight, Keith},
  journal={Journal of the Royal Statistical Society Series B: Statistical Methodology},
  volume={67},
  number={1},
  pages={91--108},
  year={2005},
}

@article{nouraie2025stability,
    author = {Mahdi Nouraie and Connor Smith and Samuel Muller},
    title = {{Stability Selection via Variable Decorrelation}},
    journal = {Statistics and Computing},
    volume = {36},
    number = {160},
    year = {2026}
}

@article{BJORCK1994297,
title = {{Numerics of Gram-Schmidt Orthogonalization}},
journal = {Linear Algebra and its Applications},
volume = {197-198},
pages = {297-316},
year = {1994},
author = {{\AA}. Bj{\"{o}}rck}
}

@article{a2ce9dde-7f22-3541-a877-07b4efae8445,
 author = {Jian Huang and Shuangge Ma and Cun-Hui Zhang},
 journal = {Statistica Sinica},
 number = {4},
 pages = {1603--1618},
 publisher = {Institute of Statistical Science, Academia Sinica},
 title = {{Adaptive Lasso for Sparse High-Dimensional Regression Models}},
 volume = {18},
 year = {2008}
}

@article{zou2006adaptive,
  title={The {A}daptive {L}asso and {I}ts {O}racle {P}roperties},
  author={Zou, Hui},
  journal={Journal of the American Statistical Association},
  volume={101},
  number={476},
  pages={1418--1429},
  year={2006},
}

@article{courtois2021new,
  title={New adaptive lasso approaches for variable selection in automated pharmacovigilance signal detection},
  author={Courtois, {\'E}meline and Tubert-Bitter, Pascale and Ahmed, Isma{\"\i}l},
  journal={BMC Medical Research Methodology},
  volume={21},
  number = {271},
  pages={1--17},
  year={2021},
  publisher={Springer}
}

@article{yuan2006model,
  title={Model selection and estimation in regression with grouped variables},
  author={Yuan, Ming and Lin, Yi},
  journal={Journal of the Royal Statistical Society Series B: Statistical Methodology},
  volume={68},
  number={1},
  pages={49--67},
  year={2006},
}

@article{park2008bayesian,
  title={The {B}ayesian {L}asso},
  author={Park, Trevor and Casella, George},
  journal={Journal of the American Statistical Association},
  volume={103},
  number={482},
  pages={681--686},
  year={2008},
}

@InProceedings{pmlr-v9-lorbert10b,
  title = 	 {Exploiting {C}ovariate {S}imilarity in {S}parse {R}egression via the {P}airwise {E}lastic {N}et},
  author = 	 {Lorbert, Alexander and Eis, David and Kostina, Victoria and M. Blei, David and J. Ramadge, Peter},
  booktitle = 	 {Proceedings of the Thirteenth International Conference on Artificial Intelligence and Statistics},
  pages = 	 {477--484},
  year = 	 {2010},
  volume = 	 {9},
  series = 	 {Proceedings of Machine Learning Research},
  publisher =    {PMLR},
  address = {Chia Laguna Resort, Sardinia, Italy}
}

@inproceedings{NIPS2011_33ceb07b,
author = {Grave, \'{E}douard and Obozinski, Guillaume and Bach, Francis},
title = {{Trace Lasso: a trace norm regularization for correlated designs}},
year = {2011},
publisher = {Curran Associates Inc.},
address = {Red Hook, NY},
booktitle = {Proceedings of the 24th International Conference on Neural Information Processing Systems},
pages = {2187–2195},
series = {NIPS'11}
}

@article{simon2013sparse,
  title={A {S}parse-{G}roup {L}asso},
  author={Simon, Noah and Friedman, Jerome and Hastie, Trevor and Tibshirani, Robert},
  journal={Journal of Computational and Graphical Statistics},
  volume={22},
  number={2},
  pages={231--245},
  year={2013},
}

@article{Chen_Ding_Luo_Xie_2013, 
title={Uncorrelated {L}asso}, 
volume={27}, 
number={1}, 
journal={Proceedings of the AAAI Conference on Artificial Intelligence}, 
author={Chen, Si-Bao and Ding, Chris and Luo, Bin and Xie, Ying}, 
year={2013}, 
pages={166-172} 
}

@inproceedings{NIPS2014_43feaeee,
 author = {Kong, Deguang and Fujimaki, Ryohei and Liu, Ji and Nie, Feiping and Ding, Chris},
 booktitle = {Advances in Neural Information Processing Systems},
 pages = {},
 publisher = {Curran Associates, Inc.},
 address = {Red Hook, NY},
 title = {Exclusive {F}eature {L}earning on {A}rbitrary {S}tructures via $l_{1,2}$-norm},
 volume = {27},
 year = {2014}
}

@InProceedings{pmlr-v84-takada18a,
  title = {Independently {I}nterpretable {L}asso: {A} {N}ew {R}egularizer for {S}parse {R}egression with {U}ncorrelated {V}ariables},
  author = {Takada, Masaaki and Suzuki, Taiji and Fujisawa, Hironori},
  booktitle = {Proceedings of the Twenty-First International Conference on Artificial Intelligence and Statistics},
  pages = {454--463},
  year = {2018},
  volume = {84},
  series = {Proceedings of Machine Learning Research},
  publisher = {PMLR},
address = {Playa Blanca, Lanzarote, Canary Islands, Spain}
}

@article{craig2024pretraining,
  title={Pretraining and the {L}asso},
  author={Craig, Erin and Pilanci, Mert and Menestrel, Thomas Le and Narasimhan, Balasubramanian and Rivas, Manuel A. and Gullaksen, Stein-Erik and Dehghannasiri, Roozbeh and Salzman, Julia and Taylor, Jonathan and Tibshirani, Robert},
  journal={arXiv preprint arXiv:2401.12911},
  year={2024}
}

@article{kalousis2007stability,
  title={Stability of feature selection algorithms: a study on high-dimensional spaces},
  author={Kalousis, Alexandros and Prados, Julien and Hilario, Melanie},
  journal={Knowledge and Information Systems},
  volume={12},
  number = {1},
  pages={95--116},
  year={2007},
}

@article{meinshausen2010stability,
  title={Stability selection},
  author={Nicolai Meinshausen and Peter B{\"u}hlmann},
  journal={Journal of the Royal Statistical Society Series B: Statistical Methodology},
  volume={72},
  number={4},
  pages={417--473},
  year={2010},
}

@article{nouraie2024selection,
  title={On the {S}election {S}tability of {S}tability {S}election and {I}ts {A}pplications},
  author={Nouraie, Mahdi and Muller, Samuel},
  journal={arXiv preprint arXiv:2411.09097},
  year={2024}
}

@article{faletto2022cluster,
  title={Cluster {S}tability {S}election},
  author={Faletto, Gregory and Bien, Jacob},
  journal={arXiv preprint arXiv:2201.00494},
  year={2022}
}

@article{bondell2008simultaneous,
  title={{S}imultaneous {R}egression {S}hrinkage, {V}ariable {S}election, and {S}upervised {C}lustering of {P}redictors with {OSCAR}},
  author={Bondell, Howard D. and Reich, Brian J.},
  journal={Biometrics},
  volume={64},
  number={1},
  pages={115--123},
  year={2008},
}

@ARTICLE{6840355,
  author={Zeng, Xiangrong and Figueiredo, Mário A. T.},
  journal={IEEE Signal Processing Letters}, 
  title={Decreasing Weighted Sorted  ${\ell_1}$ Regularization}, 
  year={2014},
  volume={21},
  number={10},
  pages={1240-1244}
}

@article{10.1093/bioinformatics/bty750,
    author = {Wang, Haohan and Lengerich, Benjamin J and Aragam, Bryon and Xing, Eric P},
    title = {Precision {L}asso: accounting for correlations and linear dependencies in high-dimensional genomic data},
    journal = {Bioinformatics},
    volume = {35},
    number = {7},
    pages = {1181-1187},
    year = {2018}
}

@book{bhl2011,
    author = {Peter B\"uhlmann and Sara van de Geer},
    title = {{Statistics for High-Dimensional Data: Methods, Theory and Applications}},
    publisher = {Springer},
    address   = {Heidelberg},
    year = {2011} 
}

@book{pareto1896cours,
  title = {Cours d'{\'E}conomie Politique},
  author={Pareto, Vilfredo},
  year={1896},
  publisher={F. Rouge},
  address = {Lausanne}
}

@article{friedman2010regularization,
  title={{Regularization Paths for Generalized Linear Models via Coordinate Descent}},
  author={Friedman, Jerome and Hastie, Trevor and Tibshirani, Rob},
  journal={Journal of Statistical Software},
  volume={33},
  number={1},
  pages={1--22},
  year={2010},
  publisher={NIH Public Access}
}

@article{10.1371/journal.pgen.0020006,
    author = {Lan, Hong AND Chen, Meng AND Flowers, Jessica B AND Yandell, Brian S AND Stapleton, Donnie S AND Mata, Christine M AND Mui, Eric Ton-Keen AND Flowers, Matthew T AND Schueler, Kathryn L AND Manly, Kenneth F AND Williams, Robert W AND Kendziorski, Christina AND Attie, Alan D},
    journal = {PLOS Genetics},
    publisher = {Public Library of Science},
    title = {{Combined Expression Trait Correlations and Expression Quantitative Trait Locus Mapping}},
    year = {2006},
    volume = {2},
    number = {1},
    pages = {0051-0061},
}

@article{10.1111/rssb.12095,
    author = {Song, Qifan and Liang, Faming},
    title = {{A Split-and-Merge Bayesian Variable Selection Approach for Ultrahigh Dimensional Regression}},
    journal = {Journal of the Royal Statistical Society Series B: Statistical Methodology},
    volume = {77},
    number = {5},
    pages = {947-972},
    year = {2014},
}

@article{10.1214/22-BA1351,
author = {Christian Staerk and Maria Kateri and Ioannis Ntzoufras},
title = {{A Metropolized Adaptive Subspace Algorithm for High-Dimensional Bayesian Variable Selection}},
volume = {19},
journal = {Bayesian Analysis},
number = {1},
publisher = {International Society for Bayesian Analysis},
pages = {261 -- 291},
year = {2024},
}

@article{JMLR:v25:23-0536,
  author  = {Jake A. Soloff and Rina Foygel Barber and Rebecca Willett},
  title   = {Bagging Provides Assumption-free Stability},
  journal = {Journal of Machine Learning Research},
  year    = {2024},
  volume  = {25},
  number  = {131},
  pages   = {1--35}
}

@article{zhao2006model,
  title={{On Model Selection Consistency of Lasso}},
  author={Zhao, Peng and Yu, Bin},
  journal={Journal of Machine Learning Research},
  volume={7},
  pages={2541--2563},
  year={2006}
}

@article{a1294e48-ce43-314c-867d-a032d7baf484,
 author = {Chenlei Leng and Yi Lin and Grace Wahba},
 journal = {Statistica Sinica},
 number = {4},
 pages = {1273--1284},
 publisher = {Institute of Statistical Science, Academia Sinica},
 title = {A NOTE ON THE {L}ASSO AND RELATED PROCEDURES IN MODEL SELECTION},
 volume = {16},
 year = {2006}
}

@article{10.1111/j.1467-9868.2008.00674.x,
    author = {Fan, Jianqing and Lv, Jinchi},
    title = {Sure Independence Screening for Ultrahigh Dimensional Feature Space},
    journal = {Journal of the Royal Statistical Society Series B: Statistical Methodology},
    volume = {70},
    number = {5},
    pages = {849-911},
    year = {2008}
}

@article{hoerl1970ridge,
  title={Ridge regression: Biased estimation for nonorthogonal problems},
  author={Hoerl, Arthur E and Kennard, Robert W},
  journal={Technometrics},
  volume={12},
  number={1},
  pages={55--67},
  year={1970},
  publisher={Taylor \& Francis}
}

@article{chatterjee2025univariate,
  title={Univariate-Guided Sparse Regression},
  author={Chatterjee, Sourav and Hastie, Trevor and Tibshirani, Robert},
  journal={arXiv preprint arXiv:2501.18360},
  year={2025}
}

@Article{ncvreg,
    author = {Patrick Breheny and Jian Huang},
    title = {Coordinate descent algorithms for nonconvex penalized regression,
	with applications to biological feature selection},
    journal = {Annals of Applied Statistics},
    year = {2011},
    volume = {5},
    pages = {232--253},
    number = {1}
  }

@Manual{simon2018package,
    title = {SGL: Fit a GLM (or Cox Model) with a Combination of Lasso and Group
Lasso Regularization},
    author = {Noah Simon and Jerome Friedman and Trevor Hastie and Rob Tibshirani},
    year = {2019},
    note = {R package version 1.3}
  }

@article{Fan01122001,
author = {Jianqing Fan and Runze Li},
title = {Variable Selection via Nonconcave Penalized Likelihood and its Oracle Properties},
journal = {Journal of the American Statistical Association},
volume = {96},
number = {456},
pages = {1348--1360},
year = {2001},
publisher = {ASA Website}
}

@article{MCP,
    author = {Cun-Hui Zhang},
    title = {Nearly unbiased variable selection under minimax concave penalty},
    journal = {The Annals of Statistics},
    volume = {38},
    number = {2},
    pages = {894--942},
    year = {2010}
}

@article{10.5555/2567709.2567772,
author = {Sun, Wei and Wang, Junhui and Fang, Yixin},
title = {Consistent selection of tuning parameters via variable selection stability},
year = {2013},
volume = {14},
number = {1},
journal = {Journal of Machine Learning Research},
pages = {3419–3440}
}

@article{feser2023sparse,
  title={{Sparse-group SLOPE: adaptive bi-level selection with FDR-control}},
  author={Feser, Fabio and Evangelou, Marina},
  journal={arXiv preprint arXiv:2305.09467},
  year={2023}
}

@article{https://doi.org/10.1111/j.1751-5823.2010.00108.x,
author = {Müller, Samuel and Welsh, Alan H.},
title = {On Model Selection Curves},
journal = {International Statistical Review},
volume = {78},
number = {2},
pages = {240-256},
year = {2010}
}

@article{tikhonov1963solution,
  title={Solution of incorrectly formulated problems and the regularization method},
  author={Tikhonov, Andrei N},
  journal={Soviet Mathematics Doklady},
  volume={4},
  pages={1035--1038},
  year={1963}
}

@article{10.1111/rssb.12427,
    author = {Wang, Xiangyu and Leng, Chenlei and Boot, Tom},
    title = {{Wang and Leng (2016), High-Dimensional Ordinary Least-Squares Projection for Screening Variables, Journal of The Royal Statistical Society Series B, 78, 589–611}},
    journal = {Journal of the Royal Statistical Society Series B: Statistical Methodology},
    volume = {83},
    number = {4},
    pages = {880-881},
    year = {2021}
}

\end{document}